\documentclass[useAMS,usenatbib,usegraphicx]{mn2e}
\usepackage{amssymb}
\usepackage{times}

\title{Galaxy number counts -- VI.\ An $H$-band survey of the Herschel Deep Field}
\author[Metcalfe, Shanks, Weilbacher, et al.]
{N.~Metcalfe$^1$\thanks{Email: nigel.metcalfe@durham.ac.uk}, T.~Shanks$^1$, P.M.~Weilbacher$^1$ \thanks{Present address: Astrophysikalisches Institut Potsdam, An der Sternwarte 16,D-14482 Potsdam, Germany}, H.J.~McCracken$^1$\thanks{Present address: Institut d'astrophysique de Paris, UMR 7095, 98 bis Bvd Arago, 75014 Paris, France/Observatoire de Paris, LERMA,, 61 Avenue de l'Observatoire, 75014 Paris, France}
,\newauthor R.~Fong$^1$, \& D.~Thompson$^2$
\\
$^1$Physics Department, University of Durham, South Road, Durham DH1 3LE\\
$^2$Astronomy Department, Caltech, Pasadena CA 91125, USA\\
}
\begin{document}

\date{Received 2005; Accepted 2005.}
%      in original form 2005 ???? ??

\pagerange{\pageref{firstpage}--\pageref{lastpage}}
\pubyear{2005}

\maketitle

\label{firstpage}

\begin{abstract}
We present $H$-band infra-red galaxy data to a $3\sigma$ limit of
$H\sim22.9$ and optical-infra-red colours of galaxies on the William
Herschel Deep Field (WHDF).  These data were taken from a $7'\times 7'$ area
observed for 14 hours with the $\Omega$ Prime camera on the 3.5-m Calar
Alto telescope. We also present counts derived from the HDF-S NICMOS camera
to the limit of $H\sim29$\,mag over a $0.'95\times0.'95$ area. Following previous
papers, we derive $H$-band number counts, colour-magnitude diagrams and
colour histograms for the whole $H$ selected sample. We review our Pure
Luminosity Evolution (PLE) galaxy count models based on the spectral
synthesis models of Bruzual \& Charlot. We find that our previously
assumed forms for the luminosity function agree well with those recently
derived from 2dFGRS/2MASS at $B$ and $K$, except that the 2dFGRS $K$ LF
has an unexpectedly flat slope which, if correct, could affect our
interpretation of the faintest $H$ and $K$ counts.

We find that these PLE models give an excellent fit to the WHDF $H$
band count data to $H<22.5$ and HDF count data to $H<28$. However,
if we use the flat 2dFGRS/2MASS near infra-red (NIR) LF then the predicted count is too flat
at $H>21$. We  confirm that PLE models that assume a Salpeter IMF for
early-type galaxies overestimate the average galaxy redshift in $K<20$
galaxy redshift surveys. Models that assume a steep $x=3$ IMF continue
to give better agreement with the $N(z)$ data than even models based on
a Scalo IMF, although they do show an unobserved peak in $B-H$ and $I-H$
colour distributions at faint $H$ magnitudes corresponding to $z>1$
early-type galaxies. But this feature may simply reflect a larger
scatter in optical-infra-red colours than in the optical $B-R$ colour of
early-type galaxies at this redshift. This scatter is obvious in
optical-IR colour-colour diagrams and may be explained by on-going
star-formation in an intermediate sub-population of early-type galaxies.
The numbers of EROs detected are a factor of 2--3 lower than predicted by
the early-type models that assume the Salpeter IMF and in better
agreement with those that assume the $x=3$ IMF. The tight sequence of
early-type galaxies also shows a sub-class which is simultaneously
redder in infrared bands and bluer in the bluer bands than the
classical, passive early-type galaxy; this sub-class appears at
relatively low redshifts and may constitute an intermediate age, early-type 
population. Finally, we have
also detected a candidate $z>1$ galaxy cluster using our panoramic
$H$-band observations of the WHDF.

\end{abstract}

%%%%%%%%%%%%%%%%%%%%%%%%%%%%%%%%%%%%%%%%%%%%%%%%%%%%%%%%%%%%%%%%%%%%%%%%%%%%%%%%

\section{Introduction}
In five previous papers, \citet[][hereafter Paper I]{Paper1},
\citet[][hereafter Paper II]{Paper2}, \citet[][hereafter Paper
III]{Paper3}, \citet[][hereafter Paper IV]{Paper4} and \citet[][hereafter
Paper V]{Paper5} we used photographic and CCD data to study the form
of the galaxy number-magnitude relation at both optical ($B\sim27.5$)
and infra-red ($K\sim20$) wavelengths on a $7'\times 7'$ field known as
the Herschel Deep Field (WHDF).

In this paper we present the results from a deep 
$H$-band infrared survey of the
William Herschel Deep Field (WHDF). We have imaged the entire 50
arcmin$^2$ area of the WHDF using the large format (1024x1024) Hawaii
Rockwell array in the $\Omega$ Prime camera on the Calar Alto 3.5-m telescope.
Our $\sim14$ hrs of $H$-band data reach $H_{vega}(3\sigma)\sim22.9$, effectively 
$\sim2.5$ magnitudes fainter than our previous UKIRT $K$-band data 
(Paper IV). We have also re-imaged the area in $K$, for $\sim1$\,hr, in order 
to compare with the UKIRT data.

The infra-red has the advantage of being sensitive to the underlying stellar 
mass, and much less affected by star formation history than 
optical wavelengths. Recent studies (e.g.~\citealt{CDM+02}) have suggested 
that Pure Luminosity Evolution (PLE) models can provide a good description
of infra-red galaxy counts and redshift distributions. Here we examine
this question with our own PLE models and our multi-wavelength data.

The paper is organised as follows: Sections~\ref{sect:theobs} and
\ref{sect:data} deal with the observations and our data reduction
techniques while Sects~\ref{sect:calib} and \ref{sect:anal} addresses
photometric calibration and image analysis. In Sect.~\ref{sect:models}
we discuss the parameters and procedures used to create our galaxy
evolution models, especially in the light of recent work on galaxy
evolution functions, before Sect.~\ref{sect:results} presents our results
in terms of galaxy number counts, galaxy colours, and extremely red
objects (EROs), as well as indications for a new high-$z$ galaxy
cluster. Sect.~\ref{sect:conc} finally summarises our main results.

%%%%%%%%%%%%%%%%%%%%%%%%%%%%%%%%%%%%%%%%%%%%%%%%%%%%%%%%%%%%%%%%%%%%%%%%%%%%%%%%

\setcounter{table}{0}
\begin{table*}
\begin{minipage}{170mm}
\caption{Observational details for the Calar--Alto and HDFS images. HDFS
magnitudes have been converted to $H$ assuming $H=F160W_{AB}-1.3$}
\begin{tabular}{@{}lcccccccc@{}}
\hline
Frame&Area&Effective&FWHM& {$3\sigma$ limit 
\footnote{Magnitude is the {\it total} magnitude of an unresolved object which would give a $3\sigma$ detection inside an aperture with the minimum radius.}}
&{$1\sigma$ isophote
\footnote{Inside 1 arcsec$^2$}}&Min. Kron&Multiplying&Correction to\cr
&(deg$^2)$&exp. (hrs)&($''$)&(mag)&(mag/$\sq\arcmin$)&radius ($''$)&factor&total (mag)\cr
\hline
CA $H$&$1.33\times10^{-2}$&14.25&0.9&22.9&23.90&0.90&1.45&0.29\cr
CA $K^{\prime}$&$1.31\times10^{-2}$&0.9&0.9&20.2&21.25&0.95&1.50&0.26\cr
CA/UKIRT $K$&$1.35\times10^{-2}$&-&1.2&20.7&21.90&1.10&1.50&0.26\cr
HDFS $F160W$&$2.5\times10^{-4}$&$\sim36$&$\sim0.3$&27.5&28.7&0.50&2.0&0.11\cr
\hline
\end{tabular}
\end{minipage}
\label{tab:details}
\end{table*}

\section{The Observations}\label{sect:theobs}
\subsection{Calar Alto}
Our data were taken during a five night observing run in August 1997 at
the f/3.5 prime focus of the Calar Alto 3.5-m telescope in the Sierra
de Los Filabres in Andalucia, southern Spain.  The $\Omega$ Prime
infra-red camera (\cite{Bizenberger98}) contained a $1024\times1024$ 
pixel HgCdTe Rockwell
HAWAII array, with a scale of $0.396''$/pixel, resulting in a field of
$6.8'\times6.8'$, ideally matched to our optical WHDF (Paper V).

Observing conditions were generally good, with `seeing' of under $1''$
on all five nights, and only one night significantly affected by cloud.

Our primary objective was to image the WHDF as deeply as
possible. Although our previous data were taken in the $K$-band (Paper
IV), as $H-K$ is only weakly dependent on $K$ (for galaxies in the range
$0.5<z<2$), our scientific objectives could be satisfied by observations
in either.  $\Omega$ Prime was designed without a cold pupil stop,
which effectively limits observations to wavelengths shortward of
$2.2\mu m$ where thermal radiation from the telescope structure is
not a problem. Even in the Calar Alto $K^{\prime}$ filter, designed
to cut off the redward end of the standard $K$-band and hence reduce
the thermal background, only $3$ second integrations were possible
before $\Omega$ Prime saturated, with a measured sky brightness of
$\sim 11.5$\,mag\,arcsec$^2$. The measured background in $H$ was $\sim
13.5$\,mag\,arcsec$^{-2}$, with $8$ second integrations being possible.
Calculations performed on test frames taken at the start of the first
night clearly showed that the $H$-band held the advantage in terms of
signal-to-noise, and it was this which dictated our final choice of
filter for our deep exposure.  Only on the first night did we observe
the WHDF in $K^{\prime}$, in order to compare with our previous UKIRT
$K$-band data (Paper IV).

We adopted a `double correlated read' readout mode where the array is
read-out twice, one at the beginning of the observations and once at the
end, and the difference signal recorded.  These were stacked in batches
of 10 before being written out. This is a compromise between observing
efficiency (time is lost on each write-out) and the need to sample
adequately variations in the background sky, which changes rapidly in
the infra-red. A complex dithering pattern on the sky, with shifts up
to $30''$, was applied to the exposures. Our total on-sky integration
times amounted to 14.25 hrs in $H$ and 54 mins in $K^{\prime}$.

\subsection{The Hubble Deep Field}
In addition to our Calar Alto data, we have also made our own analysis
of the Hubble Deep Field South F160W NICMOS image. This is a $\sim36$
hour exposure with a scale of 0.075''/pixel. We adopt the preliminary
revised $F160W_{AB}$ zero-point of 22.77. Due to dithering pattern used
in the original observations, the processed image provided by STSCI
has areas of lower signal-to-noise around the periphery. We therefore
trimmed these regions from the image to leave a reasonably uniform area
of 0.90\,$\sq\arcmin$, a 30 per cent drop from the full field.

Space-based $H$-band observations have an important advantage over
ground-based ones: because HST is above the Earth's atmosphere, the
absence of night-sky OH lines in this wavelength range means the
background is {\it much lower\/} than observed from ground-based
telescopes.  On-orbit sky brightnesses measured with NICMOS are
typically $\sim24$\,mag\,arcsec$^{-2}$, compared with the $\sim
13.5$\,mag\,arcsec$^{-2}$ from $\Omega$ Prime.

%%%%%%%%%%%%%%%%%%%%%%%%%%%%%%%%%%%%%%%%%%%%%%%%%%%%%%%%%%%%%%%%%%%%%%%%%%%%%%%%

\section{Data Reduction}\label{sect:data}
\subsection{Calar Alto $H$-band}

The data from the array is read out in four separate quadrants. For both the 
$H$ and $K^{\prime}$ bands the data reduction was complicated by the
fact that one of the four quadrants clearly showed non-linear behaviour, with 
the sky level (and object magnitudes) on this part of the chip being 
relatively lower for higher overall sky counts. Not only did this complicate
the stitching together of the dithered exposures, but also meant that the
flat-field varied with signal level. The various procedures we finally adopted
were chosen to give the best agreement with the 2-MASS stellar magnitudes 
on the field. The corrections applied to the rogue quadrant affected the 
magnitudes at the 10--15 per cent level.

The nature of the NIR sky background, and the lack of a cold stop
at Calar Alto, meant a different 
reduction procedure was needed from that described for the
UKIRT $K$-band data in Paper IV. Particularly at shorter
wavelengths ($\sim 1.5\mu$m), the NIR sky background is
dominated by many intense, narrow and highly variable OH airglow lines,
unlike at longer wavelengths (such as the range covered by our
$K^{\prime}$ filter) where thermal emissions from the telescope and
atmosphere are more important; these are expected to change over
much longer time-scales. Qualitatively, this is what we find in our
observations; if the data reduction procedure in Paper IV was
applied to the $H$-data, large background variations are observed
across the array, which vary in intensity and position from frame
to frame. Furthermore as part of a recent design study for the next
generation instrument, B.~Rauscher (priv.~comm.) has carried out an
independent, quantitative analysis of the sky variations as measured from
this $\Omega$ Prime data. He concludes that OH airglow, changing on
time-scales of $\sim$ 1.5 minutes is responsible for the observed 0.5 per cent
variation in the sky background. As well as these rapid variations,
the sky background changed gradually by up to a factor of three during
the five nights.

Such large sky variations meant the non-linearity in the rogue quadrant
was a severe problem for the $H$-band data. After several unsatisfactory
attempts, we eventually adopted the following reduction procedure,
which, as we show in the next section, produces good agreement with
2MASS measurements on the field (although, of course, this comparison is 
only possible for the brighter stars on the field). All the reductions 
were performed using STARLINK software.

First a 'flat' frame is constructed from the dome flat fields by
subtracting the exposures of equal length taken with the shutter open
and closed. This should remove any thermal signal coming from within
the camera. The resulting frame agrees reasonably ($\sim$2 per cent level) with
those constructed by subtracting data frames with differing background
levels from one another, suggesting that (a) our sky is really flat and
(b) the large variations in background are due to changes in sky and
not thermal signal from within the telescope.

We then grouped our data frames from all the nights into ~10 batches of 
similar background levels (a small number of frames taken on the cloudiest 
night were not used). Each batch was reduced independently as follows:
The 'flat' frame is scaled and added to or subtracted from the data frames in
order to ensure all the data frames have the same mean background level. 
At the same time a small ($\sim1$ per cent), background level dependent scaling 
factor 
is applied to the rogue quadrant to account for non-linearity.
All the frames in the batch are then median combined to produce a master
'background' frame, which is then subtracted from each frame in turn. This
procedure was found to be the only way of producing a 'background' frame
free of objects. The background-subtracted frames were then flat-fielded using
a version of our 'flat' normalised to 1. Finally, these frames were spatially 
matched, residual background variations removed by fitting a 2D 3rd order 
polynomial, and added together (with a 4 sigma cut to take out hot pixels).
The individual batches are then spatially aligned and added together to
produce the final data frame.

\subsection{Calar Alto $K$-band}

The $K^{\prime}$ background levels varied much less than in $H$. However,
the dome flat did not agree well with the sky frame found by subtracting
data frames. This is probably due to the fact that the background level in 
$K$ was very high for the data frames (and much lower for the dome flats).
As a result the flat was formed by median combining many sky frames found
from subtracting independent data frame pairs. As with $H$, the data was
split into (3) batches of similar background level and each batch reduced
independently. Once again a master background frame is calculated for 
each batch and subtracted from all the frames. These are then flat-fielded, 
re-aligned and recombined into a single image (using $5\sigma$ clipping to 
remove hot pixels and other defects). 

\begin{figure}
\begin{center}
\includegraphics[width=3.25in]{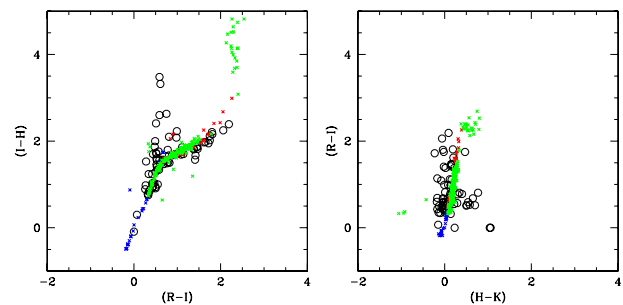}
\caption{$R-I$:$I-H$ (left panel) and $H-K$:$R-I$ (right panel)
colour-colour diagram for stars on our WHT deep field (open circles)
compared with the standard star photometry of \protect\cite{HLL+01}, \protect\cite{Legg88} and \protect\cite{Dahn02} (crosses).}
\label{fig:stars}
\end{center}
\end{figure}

\begin{figure}
\begin{center}
\includegraphics[width=3.25in]{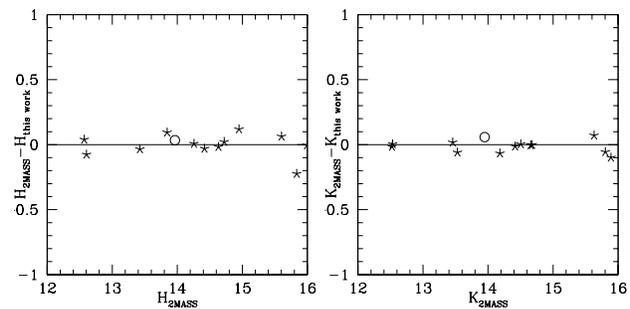}
\caption{$H$-band (left panel) and $K$-band (right panel)
comparison between our magnitudes and those for 2MASS sources
on the WHT deep field. The one galaxy is marked with an open circle. The
other objects are all stellar.}
\label{fig:2mass}
\end{center}
\end{figure}

%%%%%%%%%%%%%%%%%%%%%%%%%%%%%%%%%%%%%%%%%%%%%%%%%%%%%%%%%%%%%%%%%%%%%%%%%%%%%%%%

\section{Calibration}\label{sect:calib}
Calibration of the Calar Alto data was based on standard star observations
made on 4 of the 5 nights. Magnitudes were measured in small apertures
and extrapolated to 'total'. Standards were taken from the UKIRT faint
standards list, supplemented by the data of \cite{Hunt98}. Each star
was observed once in each quadrant, but all measurements from the rogue
quadrant were ignored.
Despite the occasional presence of cirrus, there was little difference
between the four nights on which standards were taken.  To monitor
relative conditions (and the effect of airmass) we tracked stellar
magnitudes off all the data frames throughout each night.  Agreement was
good, even on the nights with cirrus, with all the individual exposures
finally used showing zero-points within $0.1$\,mag of each other.  For stars
of good signal-to-noise, the rms magnitude over all frames on all nights
was only $0.04$\,mag. The zero-points of the final stacked frames are
corrected for these variations.

\begin{figure}
\begin{center}
\includegraphics[width=3.3in]{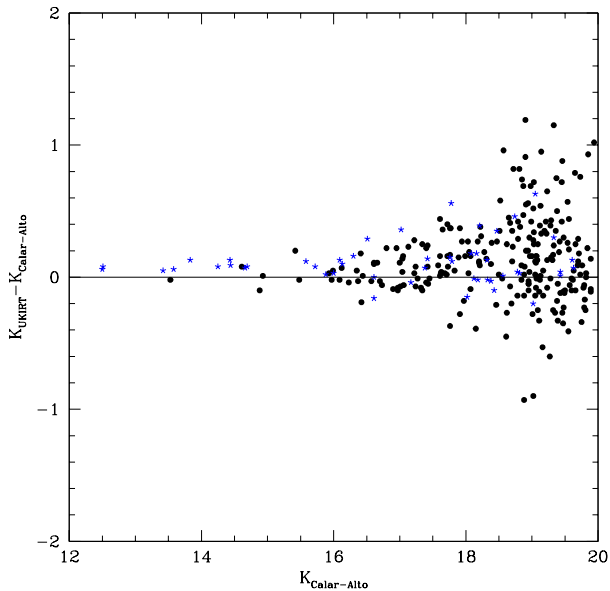}
\caption{A comparison of galaxy magnitudes from our UKIRT data (Paper IV)
and this work. There is a slight offset in that the sense that the UKIRT
data are $\sim0.06$\,mag fainter. We have ignored any colour term between
$K^{\prime}_\mathrm{CA}$ and $K_\mathrm{UKIRT}$.}
\label{fig:magcomp}
\end{center}
\end{figure}

The rms scatter about the mean offset between instrumental and standard
magnitudes was $\pm0.04$\,mag in $H$ and $\pm0.03$\,mag in $K$-prime, similar
to that between the data frames. We have neglected any colour term
between $K$-prime and $K$.  Fig \ref{fig:stars} shows the $R-I$:$I-H$
and $H-K$:$R-I$ diagrams for stars on our WHDF $H$-band  and $K$-band frames
compared with stellar photometry from \cite{Legg88}, for late-type
stars, and from the UKIRT Fundamental Standards List \citep{HLL+01} for
earlier spectral types, and \cite{Dahn02} from cool dwarfs.  Agreement is
reasonable in both cases.

We have also been able to compare our brighter stellar magnitudes with 
the available 
data from the 2MASS point source catalogue (Fig \ref{fig:2mass}). 
Ignoring the brightest star, which is saturated, we find
excellent agreement; for $H$, $H_\mathrm{2MASS}-H_\mathrm{this\ work}
= -0.00\pm0.09$ and for the $K$-band, $K_\mathrm{2MASS}-K_\mathrm{this\
work} = -0.01\pm0.05$. Note that the faintest three stars all have 2MASS
errors between 0.1 and 0.25 mags.

%%%%%%%%%%%%%%%%%%%%%%%%%%%%%%%%%%%%%%%%%%%%%%%%%%%%%%%%%%%%%%%%%%%%%%%%%%%%%%%%

\section{Image analysis}\label{sect:anal}
Our image analysis techniques have been well documented elsewhere (Papers
II,III,IV, \& V). In brief, the background sky is removed using a 2D
polynomial fit. A first pass is then made over the data using an isophotal
object detection routine to a magnitude limit much fainter than that of a
$3\sigma$ detection. Objects so detected are then removed from the frame
and replaced by a local sky value (plus appropriate noise). The resulting
image is heavily smoothed and subtracted from the original. The isophotal
object detection is then repeated on this flat-background frame. These
detections are then input to a Kron-type aperture magnitude routine from
which our final magnitudes are derived. Importantly, our Kron-radii are
not allowed to become smaller than that for an unresolved image. Kron
magnitudes require a correction to `total', which ideally is independent
of profile shape, but is dependent on the multiplying factor used to
calculate the Kron radius. As in our previous work, we adopt an unusually
small factor which results in a significant correction to `total', 
of $\sim0.3$ mag, but
does reduce the contaminating effect of close neighbours. Even so, it
is necessary to `clean' such objects. Table \ref{tab:details} lists the
parameters of all our final data frames.

Our WHDF optical-infrared and $H-K$ colours are measured in
fixed $1.5''$ radius apertures. Astrometry was provided by matching to the
USNO catalogue, using the STARLINK GAIA package.

We are in the fortunate position of being able to compare our Calar-Alto
$K'$-band magnitude with those from our independent UKIRT observations
(Paper VI), which cover the same area and are very similar in terms
of signal-to-noise. Fig. \ref{fig:magcomp} shows a plot of magnitudes
for stars and galaxies in common to both frames. The agreement is reasonably
good right down to the limit of the photometry. For $K<16$ we find $\Delta
K_\mathrm{CA-UKIRT} = -0.06\pm0.06$. At $K\sim18.5$, the noise has
increased to $\sim\pm0.3$mag.

In order to improve the signal-to-noise, image analysis was run on
the stacked $K$-band frames from UKIRT and Calar Alto. The UKIRT data
is slightly deeper, but the Calar Alto data has better image quality,
so both images were given equal weight in the stack. Unless otherwise
stated, in the rest of the paper $K$ magnitudes and $H-K$ colours refer
to those from this combined data-set.

The HDF NICMOS data was analysed in similar fashion to the Calar Alto data, 
except that the higher resolution meant that it was often necessary to 
recombine images which had been artificially split by the software into several 
parts. Such images were identified by visual inspection of the data.
A similar problem affected the HDF optical data (see Paper V), although 
the problem is not as severe in the NICMOS frames, due in part to the worse 
image quality and also due to the more regular morphology of galaxies at 
longer wavelengths.

\subsection{Star/galaxy separation}
The star-galaxy separation used on the ground-based data is that described for
the WHDF in Paper V.
Basically this was done on the WHDF $B$ image using the difference between
the total magnitude and that inside a $1''$ aperture, a technique described 
in detail in Paper II. This enabled us to separate to $B\sim24$\,mag. 
Some additional very red stars were identified from the $R$ and $I$ frames.
In $H$ this means that most objects have reliable types to $H\sim19.5$, 
slightly fainter than the limit for identifications based on the frame $H$ alone, 
which is $\sim19$\,mag. It is possible to use $H-K$ colour as a star/galaxy
separator at even fainter magnitudes -- see Fig \ref{fig:hkriplot} 
(except for the 
bluest optical colours where late type low redshift galaxies have the same 
colours are main sequence stars).
However, our relatively bright $K$-band limit restricts the usefulness of
this in our case.

%%%%%%%%%%%%%%%%%%%%%%%%%%%%%%%%%%%%%%%%%%%%%%%%%%%%%%%%%%%%%%%%%%%%%%%%%%%%%%%%

\section{Galaxy Evolution Models}\label{sect:models}
Before discussing our results in detail, we take a more considered look
at the galaxy evolution models we have used in our previous papers.
Once again we use PLE (pure luminosity evolution) models as comparison
to our observed data to demonstrate that even simple models, with no 
assumed dynamical evolution, can well explain
the observed counts \citep[see Paper V,][]{Paper5} and redshift
distributions \citep[Paper IV,][]{Paper4}.

To keep the models simple we use only two basic forms of evolution,
one for early type galaxies (E, S0, and Sab) with a characteristic
time-scale of $\tau=2.5$\,Gyr and one for late types (Sbc, Scd, and
Sdm) with $\tau=9$\,Gyr, both in the context of the Bruzual \& Charlot
(2000) evolutionary library. We generally use a cosmology with Hubble
constant of $H_0=50$\,km\,s$^{-1}$\,Mpc$^{-1}$ and a high ($q_0=0.5$)
or low ($q_0=0.05$) density, or with a flat cosmology and cosmological
constant ($\Omega_M=0.3$, $\Omega_\lambda=0.7$). The formation redshifts
for these models are $z_\mathrm{f}=9.9$, $z_\mathrm{f}=6.3$, and
$z_\mathrm{f}=7.9$, respectively. The actual cosmology used will be
referred to as appropriate.

\subsection{Luminosity functions}
These models, when taking into account the cosmology and the attenuation
due to intervening hydrogen clouds, directly predict the evolution in
colour space. Convolution of this galaxy evolution (taking $e$ and $k$
corrections together) with a type dependent luminosity function (LF)
finally gives us predictions for number counts, redshift distributions,
colour histograms, and various other observables. This means that the LF
is a critical ingredient of our modelling technique which needs to be
checked, taking into account the most recent developments in the study
of LFs from different surveys.

\begin{table}
\caption{Rest frame parameters of the luminosity function used for the
         PLE models ($H_0 = 50$ kms$^{-1}$Mpc$^{-1}$).}
\label{tab:hlf}
\centering
\begin{tabular}{@{}lcccccc@{}}
\hline
Type & $\phi^*$ (Mpc$^{-3}$) & $\alpha$ & $M_H^*$ & $M_K^*$ & $R$-$H$ & $R$-$K$ \\
\hline
E/S0 & $9.27\times10^{-4}$  & -0.7  &  -24.85 &  -24.92 & 2.41  & 2.48  \\ % B-H=4.00
Sab  & $4.63\times10^{-4}$  & -0.7  &  -24.27 &  -24.78 & 2.01  & 2.52  \\ % B-H=3.39
Sbc  & $6.20\times10^{-4}$  & -1.1  &  -24.41 &  -24.83 & 2.03  & 2.45  \\ % B-H=3.19
Scd  & $2.73\times10^{-4}$  & -1.5  &  -24.28 &  -24.34 & 2.07  & 2.13  \\ % B-H=2.91
Sdm  & $1.36\times10^{-4}$  & -1.5  &  -23.70 &  -23.71 & 1.57  & 1.58  \\ % B-H=2.32
\hline
\end{tabular}
\end{table}

\begin{figure}
\begin{center}
\includegraphics[width=6cm,angle=270]{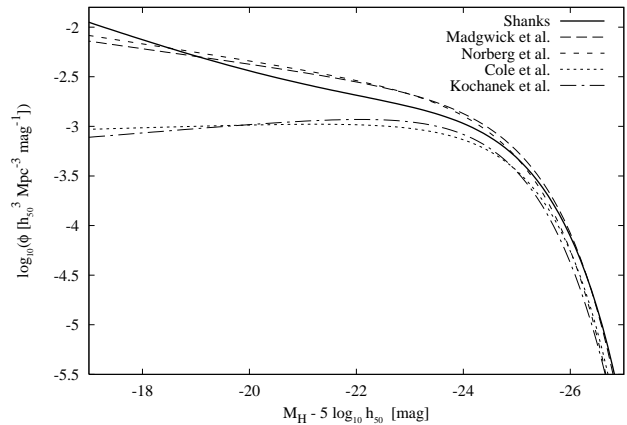}
\caption{$H$-band luminosity functions. The bold solid
  line is our LF from Table~\ref{tab:hlf}, the long and short dashed lines
  are the LFs derived in the optical by the 2dFGRS project, and dotted
  and dashed dotted lines are the LFs derived in the NIR from
  2MASS photometry. 
}
\label{fig:totalLFs}
\end{center}
\end{figure}

The parameters of our LF (see Table~\ref{tab:hlf}) are derived from the
ones we previously used in the optical regime (see Table 13 of Paper
V). We adapted them to the NIR through the mean colours of galaxies of
each type, as given in Table~\ref{tab:hlf}. For Paper V we checked our LF
with early results of the total LF from the 2dF Galaxy Redshift Survey
and found good agreement.  Now we can also check our type dependent
LF with that of the 2dFGRS project \citep{MLB+02} in the blue $b_J$
band. As we are interested mainly in the performance of the LFs in the
NIR regime, we convert them to the $H$-band using the mean galaxy colours
for each of our 5 galaxy types.  For the total LFs we density-weight
the optical and NIR luminosities to form a total LF of
all galaxies. The result is shown in Fig.~\ref{fig:totalLFs}. It is
apparent that the LFs derived by the 2dFGRS team \citep{MLB+02,NCB+02}
agree with ours to magnitudes at least as faint as $M_H^\star-6$. As
the relevant SDSS publications find a good agreement between their LF and
those derived by the 2dFGRS team, we skip the comparison with their data.

We also compare our
LF to that derived from 2MASS data, especially the analyses by
\citet{CNB+01} and \citet{KPF+01} that are widely used in the
literature. Even when looking at the parameters of these LFs it
seems that they are at odds with the ones derived in the optical:
faint end slopes of $\alpha_\mathrm{NIR}\approx-0.9$ do not agree
with $\alpha_\mathrm{opt}\approx-1.2$ as derived in the optical. When
converting to the NIR via mean galaxy colours, the faint
end slope of the 2dFGRS LFs is much steeper. This becomes clear when
looking at the plot in Fig.~\ref{fig:totalLFs} where the faint end of
the NIR derived LFs has an order of magnitude less galaxies
than the optically derived luminosity functions. Additionally, the
NIR LFs also show a lower total galaxy density.
A possible explanation for this discrepancy is already given by
\citet{CNB+01}: the 2MASS catalogue could be biased against faint
galaxies. It is known that 2MASS misses low-surface brightness galaxies
that are within the nominal magnitude limit of the survey \citep[see
e.g.\-][]{BMK+03}.

As in this paper we are interested in more accurate descriptions of
faint galaxies and in studying galaxy properties over the whole optical to
NIR wavelength range, we prefer to use the luminosity function
as presented in Table~\ref{tab:hlf} for our models over newer, NIR
derived LFs.

\subsection{Initial Mass Function}\label{sect:IMF}

\begin{figure}
\begin{center}
\includegraphics[width=\hsize]{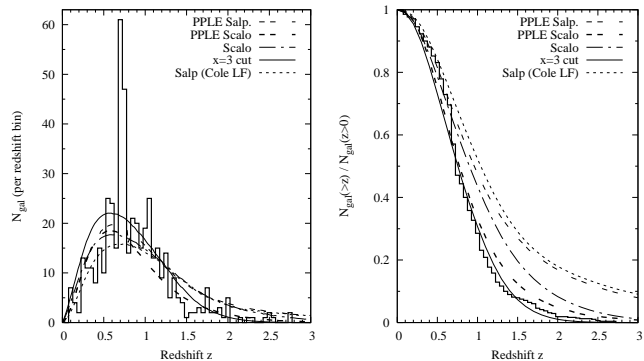}
\caption{Observed and modelled redshift distributions presented as
         differential (left) and fractional cumulative redshift
         distributions (right) for the magnitude range $K<20$\,mag.
         Shown are the 480 galaxies of the K20 survey of which 417 have
         spectroscopic redshifts (bold solid histogram).  Models are:
         PPLE models of the K20 survey team with Salpeter and Scalo IMF
         (thin and bold dashed, resp.), Scalo and $x=3$ cut IMF with our
         LF (long-short dashed and solid lines), and a model using the
         Cole LF and a Salpeter IMF (short dashed).}
\label{fig:N_of_z}
\end{center}
\end{figure}

\begin{figure}
\begin{center}
\includegraphics[width=\hsize]{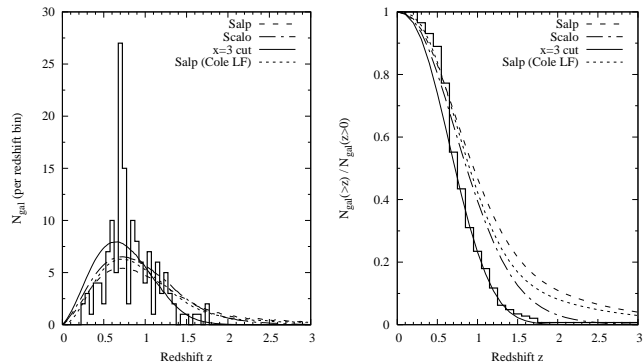}
\caption{Same as Fig.~\ref{fig:N_of_z} but just for the magnitude range
         $18<K<19$\,mag. The PPLE models of the K20 team are not available
         for this magnitude range, instead we show our model with Salpeter
         IMF.}
\label{fig:N_of_z_18to19}
\end{center}
\end{figure}

In the past, our model predictions could only be reconciled with both
the observed NIR galaxy redshift distributions $N(z)$ and the
observed galaxy number counts using a non-standard initial mass function
(IMF) for early type galaxies with a slope of $x=3$ and a low-mass
cutoff at 0.5\,M$_\odot$ while late types are modelled with a
\cite{Sal55} IMF. Using only standard Salpeter (x=1.35) or \cite{Sca86}
(x=2.5 at high stellar masses) IMFs for all galaxy types would predict
more high redshift galaxies than are observed (Paper IV). While
some local analyses of early type galaxies do not find evidence for a
steep IMF, the detailed investigation by \cite{VPB+97} of several
elliptical and lenticular galaxies using both broad-band colours and
spectral indices confirms that a steep IMF is a possibility in at least
a fraction of early-type galaxies at $z=0$.

Apart from surveys that determine the redshift distributions through
photometric methods complete to some limit, 
the only recent spectroscopic, infra-red redshift survey is the the 
K20 survey \citep{CDM+02}. They determined the redshifts of 
480 galaxies down to a limit of $K_s=20$
in an area of 52\,$\sq\arcmin$ to high completeness. Here, we use their
redshift distribution as a test for our models. While the rest of this
paper is mainly based on the $H$-band data, we have to use the $K$-band
for this particular task.

First, we compare our PLE models with \cite{Sal55} and $x=3$ cut
IMFs to the data and the ``PPLE'' models presented by the K20 team in
\citet{CPM+02}.  In Fig.~\ref{fig:N_of_z} we show the observed redshift
distribution of \citet{CPM+02} together with the predictions of models
with different IMFs in the magnitude range to the survey limit of
$K<20$\,mag. To ease comparison with the Cimatti paper, we use the
``concordance'' cosmology with $H_0=70$\,km\,s$^{-1}$\,Mpc$^{-1}$,
$\Omega_M=0.3$, and $\Omega_\Lambda=0.7$ here.  We also correct the
models for the effect that the apparent magnitude of the bulk of the
K20 sample is slightly underestimated, by computing the $N(z)$ to depths
of 19.75\,mag for early and 19.9\,mag for late type galaxies instead of
the nominal 20\,mag.

The comparison in Fig.~\ref{fig:N_of_z} presents the $N(z)$
distributions in two different ways. On the left we show the histogram
of K20 spectroscopic redshifts (supplemented by a few photometric
redshifts to be complete to $K=20$). As can be seen, the PPLE Scalo
model that was selected as best fit model by Cimatti et al.\ well
represents the observed distribution while their PPLE Salpeter model
does not give a good fit and especially overpredicts galaxy numbers at
redshifts $z>1.2$. We try to get a good match to the data with three
different models\footnote{Models with the combination of the Salpeter
IMF and our ``Shanks'' LF give worse fits than the other models, so we
do not plot them in Fig.~\ref{fig:N_of_z}}: two models both using our
luminosity function, one with Scalo IMF (like the one used in Paper IV)
and one with the $x=3$ cut IMF, and additionally a model with Salpeter
IMF and the Cole et al.\ LF. Of these three models, the Scalo and
Salpeter models again overpredict the galaxy numbers at high redshift
while the $x=3$ cut model overpredicts the numbers around $z=0.4$. On
the righthand side of Fig.~\ref{fig:N_of_z} we show an easier way to
judge the quality of the fit in the form of fractional cumulative
redshift distributions that show the number of galaxies with redshifts
higher than a given redshift. While in this representation most models
lie to the right of the data histogram, i.e.\ predict higher numbers of
galaxies at higher redshift, the $x=3$ cut model very well fits the
overall shape of the data if local variations due to clustering are
disregarded. The fit for this model is even better than for Cimatti et
al.'s Scalo model, especially at high redshifts with $z\gtrsim0.8$.

While the comparison of the $N(z)$ shape is difficult with samples
of this relatively small size because of the effect of clustering, 
which dominates
the histogram in certain redshift slices, one can also use the median
redshift as a quick measure of how well the models compare to the data.
The median survey redshift is given as $z=0.74$ for all 480 galaxies
(or $z=0.81$ disregarding the clusters at $z\sim0.7$). This as well as
the median redshifts of the different models can be read off the graph
in Fig.~\ref{fig:N_of_z}b. The PPLE Scalo model ($z_\mathrm{med}=0.78$)
and the $x=3$ cut model ($z_\mathrm{med}=0.75$) very well agree with the
median redshift of the survey while the models with all other IMFs (PPLE
Salpeter, $z_\mathrm{med}=0.98$, our Scalo model, $z_\mathrm{med}=0.90$,
and our Salpeter model with Cole LF, $z_\mathrm{med}=0.95$) do not match
the observed data.

In Paper IV we chose to use the model with $x=3$ cut IMF as our main
model after comparing redshift distributions in the magnitude range
$18<K<19$\,mag. We therefore carry out an additional test with a subset
of the K20 data in this magnitude bin. The result is presented in
Fig.~\ref{fig:N_of_z_18to19}. While it is not possible to test how 
well the models of the K20 team compare to these data, we can conclude 
that, amongst  our models, that with the $x=3$ cut IMF has the best fit. 
Again, it does not seem to be
possible to get a similarly good agreement using other IMFs like Scalo
or Salpeter, even if the Cole luminosity function is used.

\begin{figure}
\begin{center}
\includegraphics[width=\hsize]{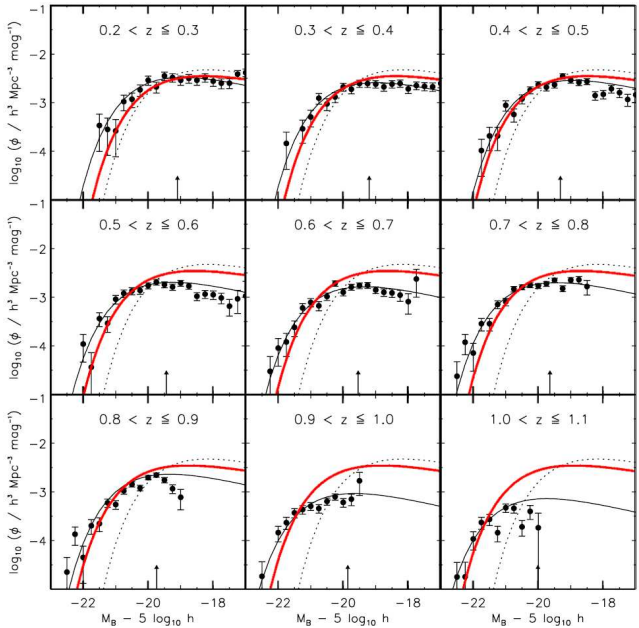}
\caption{Comparison between our predicted LF's for E/SO galaxies (bold
         solid line) with those from the COMBO-17 photometric redshifts
         of \citet{BWM+03} in the rest $B$-band. The light solid line represents
         the fit of \citet{BWM+03} to the observed LF and the  dashed line represents
         the SDSS local E/SO LF.}
\label{fig:combo}
\end{center}
\end{figure}

Finally, we compare our predictions for E/SO galaxies out to $z\approx1$
with the photo-z COMBO-17 data of \citet[][see
Fig.~\ref{fig:combo}]{BWM+03}. The COMBO-17 LFs are expressed in the
rest $B$-band. The models again appear to give a reasonable fit to the
data out to $z\sim0.8$ which shows that the small amount of  luminosity
evolution in the data is well matched by the $x=3$ model. In addition,
we also find our early-type model is in good agreement  with preliminary
results from 2dF-SDSS redshift survey of Luminous Red Galaxies out
to $z\approx0.7$ (D.\ Wake, priv.\ comm.). Early results at even higher
redshift from the VIMOS/VLT Deep Survey also show virtually no
evolution for the red galaxy LF and strong luminosity evolution for the
blue galaxy LF \citep{LeF04}, both characteristics of our PLE models.

These results show that even the simple models with just two basic types
can be used to quite well interpret data in deep fields and redshift
surveys and that the models with the $x=3$ cut IMF, only slightly
steeper than the Scalo IMF at high stellar masses, agree best with the
data available to us. We, therefore, again adopt this model, with the $x=3$
cut IMF and luminosity function as in Table~\ref{tab:hlf}, as our main
model.

%%%%%%%%%%%%%%%%%%%%%%%%%%%%%%%%%%%%%%%%%%%%%%%%%%%%%%%%%%%%%%%%%%%%%%%%%%%%%%%%

\section{Results \& Discussion}\label{sect:results}

\subsection{$H$-band counts}\label{sect:Hcounts}
\begin{figure*}
\begin{center}
\includegraphics[width=\hsize]{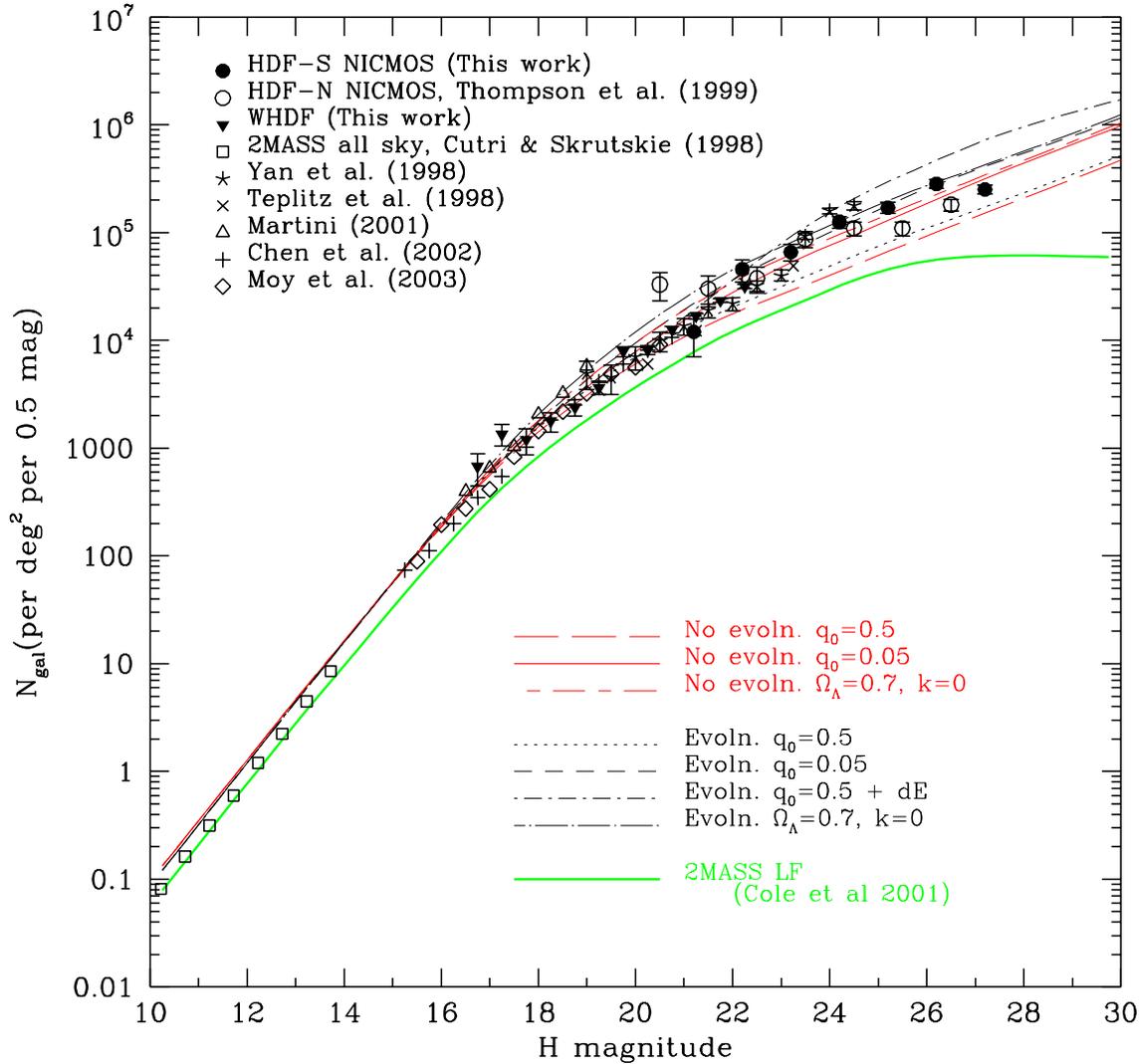}
\caption{$H$-band number-counts from the literature and from our current work.
A correction has been
  applied to the number counts of \citet{Thom99} using their supplied
  incompleteness corrections. Also shown are the predictions of the various
  models discussed in the text -- a low- and high- $q_0$ non-evolving model,
together with a non-evolving lambda model, and four evolutionary models, low
and high $q_0$, high $q_0$ with an added dwarf galaxy component and a
spatially flat lambda model. All of these assume the LF parameters in Table 2.
The 2dFGRS/2MASS model is based on the LF of  \citet{CNB+01} and assumes the same PLE 
$k+e$ corrections as the other evolutionary models for the red and blue galaxies.}
\label{fig:counts}
\end{center}
\end{figure*}

In Fig.~\ref{fig:counts} we show our differential $H$-band galaxy
number-counts measured in the WHDF and from the HDFS NICMOS frame, as
well as counts from the literature. Poisson error bars are shown where
available. There is now very good agreement between
the published datasets in the range $16<H<21$. Faintward of this the
scatter increases, but it must be remembered that these data come from
very small areas of sky, and that cosmic variance has not been included
in the error bars. The counts at bright magnitudes are the 2MASS all-sky
XSC counts above a galactic latitude of $\pm25\deg$, with a $0.15$\,mag
correction \citep{Jar00} to total magnitudes and show that the
normalisation of the models (which adopt the $B$-band luminosity function
extrapolated to $H$ using the rest-frame colours of galaxies) is
reasonable (see \citealt{Frith03} for a discussion of a local
underdensity in the infra-red counts).

Table \ref{tab:whdf counts} lists the $H$ counts from our Calar Alto data.
Table \ref{tab:hdf counts} details the HDFS NICMOS counts. To transform
from $F160W_{AB}$ to $H$ requires a brightening of 1.3\,mag (as computed
with the {\tt synphot} tool).

To allow for the extended nature of many galaxies, we only quote our
counts to $H=22.5$ rather than the $3\sigma$ limit for stellar sources
of $22.9$\,mag.

\begin{table}
\centering
\caption{$H$-band differential galaxy counts from the 14hr Calar Alto
frame.}
\begin{tabular}{@{}lcc@{}}
\hline
Magnitude & \multispan2\hss Raw N$_{gal}$\hss \cr
($H$) & (frame total) & ($\deg^{-2}$ (0.5mag)$^{-1}$)\cr
\hline

17.0-17.5 &  18 &  1349\cr
17.5-18.0 &  16 &  1199\cr
18.0-18.5 &  24 &  1798\cr
18.5-19.0 &  32 &  2398\cr
19.0-19.5 &  49 &  3672\cr
19.5-20.0 & 105 &  7868\cr
20.0-20.5 & 109 &  8167\cr
20.5-21.0 & 169 & 12663\cr
21.0-21.5 & 226 & 16934\cr
21.5-22.0 & 314 & 23528\cr
22.0-22.5 & 419 & 31396\cr

\hline
\end{tabular}

\label{tab:whdf counts}
\end{table}

\begin{table}
\centering
\caption{$F160W_{AB}$-band differential galaxy counts from the HDF South
NICMOS field.}
\begin{tabular}{@{}lcc@{}}
\hline
Magnitude & \multispan2 \hfil N$_{gal}\hfil$\cr
($F160W_{AB}$) & (frame total) & ($\deg^{-2}$ (0.5mag)$^{-1}$)\cr
\hline
23.0-24.0 &  18 &  35900\cr
24.0-25.0 &  32 &  64000\cr
25.0-26.0 &  63 & 125500\cr
26.0-27.0 &  78 & 155500\cr
27.0-28.0 & 138 & 275500\cr
28.0-29.0 & 126 & 251500\cr
\hline
\end{tabular}
\label{tab:hdf counts}
\end{table}

At brighter magnitudes our work agrees with the published counts. In
particular, our counts are close to those from the $0.4\,\sq^\circ$ area
published by \cite{CMM+02}. Faintward of $H\sim22.5$ (where all the
counts are HST based) we find our HDFS counts higher than the HDFN
counts of \cite{Thom99}, but lower than \cite{Yan98}. This may be due to
cosmic variance, or even differences in the way that the various data
reduction software cope with the tendency of faint objects to break into
sub-detections (section 5).

In the main, we consider model counts based on the LF parameters in Table 2, 
for consistency with the work in the optical count models in Paper 5. As
noted in Sect 6.1 there is reasonable agreement between this and other
more recently determined optical LFs at brighter absolute magnitudes, 
although infra-red determined LFs seem to have a flatter faint end slope. 
We shall see that this variation in slope
will cause differences in interpretation of the faintest HDF counts.

As far as comparison with the PLE models with the LF parameters in Table
2 is concerned, it is apparent that faintwards of $H\sim21$, our number
counts are higher than the predictions of the both the evolving and
non-evolving $q_0=0.5$ models; the faint end of the $K$-selected counts
has already hinted at a similar trend, as illustrated in Fig. 1 in Paper
IV. Apart from this you would be hard pressed to distinguish
between the various models. The alternative dwarf-dominated $q_0=0.5$
model proposed by \cite{Metc96} to explain the optical counts is
probably too high to fit the faintest bins of the HDF data, but could be
lowered at faint magnitudes somewhat without destroying the agreement in
the optical (Paper V) (In this ``disappearing dwarf'' model, the dwarf
population has constant star-formation rate at $z>1$ and fades rapidly
at $z<1$). The $\Lambda$ dominated cosmology gives too high a count at
intermediate magnitudes ($19<H<21$) for our relatively high
normalisation (see Paper V for a discussion of the choice of
normalisation), but would probably satisfy those who favour a lower
value. Our $q_0=0.05$ $x=3$ evolutionary PLE model reproduces the
observed number counts well.

Of course, if has been known for some time that 
the \emph{optically} selected number counts diverge from the
$q_0=0.5$ NE model, at around $B\sim20$, when the effects of
evolutionary brightening become significant. And at NIR
wavelengths we expect the morphological mix to become spiral dominated
faintwards of $K\sim20$, so it is not too surprising that the counts
should be above the predictions of the non-evolving $q_0=0.5$ model,
which, after all, fails to reproduce the number counts correctly in all
other bandpasses.  The amount of passive evolution in $H$ must still be
small by $H\sim22$, however, given that our observations are close to our
$q_0=0.05$ evolutionary model, which has been specially tuned to reduce
the amount of passive evolution. Furthermore, the angular correlation
function of this sample, discussed in \cite{McCr00}, appears also to
favour an essentially non-evolving redshift distribution at these
depths. 

Finally, Fig.~\ref{fig:counts} also shows results from modelling the
galaxy number counts with the more recent NIR LF of \citet{CNB+01}.  In the
context of our simple models using two basic evolutionary tracks, these
LFs produce less good fits to the faint H counts. Whether we use just
the total LF \citep[as given by][converted to the $H$-band]{CNB+01} with
$M_H^*=-24.73$ and $\alpha=-0.96$ and assuming an early-type $k+e$ correction,
or use the
LF split between our two evolutionary galaxy types, it is not possible
to derive a good fit to the data, irrespective of other model parameters
like IMF or characteristic time-scale $\tau$; the counts are too flat at
the faintest magnitudes (see Fig. \ref{fig:counts}.) Even if the
$\approx1.8\times$ higher normalisation, as assumed by our models, is used,
the predicted count would still be too low to fit the data via a PLE
model. We should also emphasize the importance of the normalisation in
determining how well models fit. Our normalisation is taken at
$B\approx18.5$ to avoid the issues with large scale structure at
brighter magnitudes \citep[see][]{BS04,Frith05}.

Thus if the local galaxy LF is closer to that given by our parameters in
Table 2, then the good fit of the models to the faint counts suggest
that the galaxy LF at high redshift($z\approx1$) has a slope similar to
the present day, with no need to evoke evolutionary steepening. This
means that there is no immediate detection of the steep LF at $z\ga1$
that would better match the generically steep halo mass function of CDM
models. However, if the flatter local LF of \cite{CNB+01} is taken then
evolutionary steepening of the LF may be necessary at high redshift. The
main argument against the \citet{CNB+01} 2MASS LF is that it does not seem
consistent with the steeper LFs found when  optical LFs are converted to
the H band using straightforward colour transformations of the galaxy
sub-populations. More checks of the 2MASS H band magnitude scale as used
by \cite{CNB+01} are required. If the 2MASS magnitudes are correct then
the excellent fit of our almost unevolving PLE models to the data in the
range $10<H<28$ may then have to be taken as coincidental.

In summary, if the infra-red galaxy LF we have derived from the optical LF is 
accurate then there is no need to invoke any evolution in the form
of the galaxy LF in the range $0\la z\la 2$ to fit the $H$-band counts.
The excellent fit of almost a non-evolving model throughout the range
$H<28$ mag range then may be an excellent indication that the high and low
redshift Universe may be more similar than usually expected. However, if
the local LF has a flatter slope and/or a lower normalisation then this
slowly evolving model is less consistent with the faintest $H$ counts data.

\subsection{Colours}
\begin{figure}
\begin{center}
\includegraphics[width=\hsize]{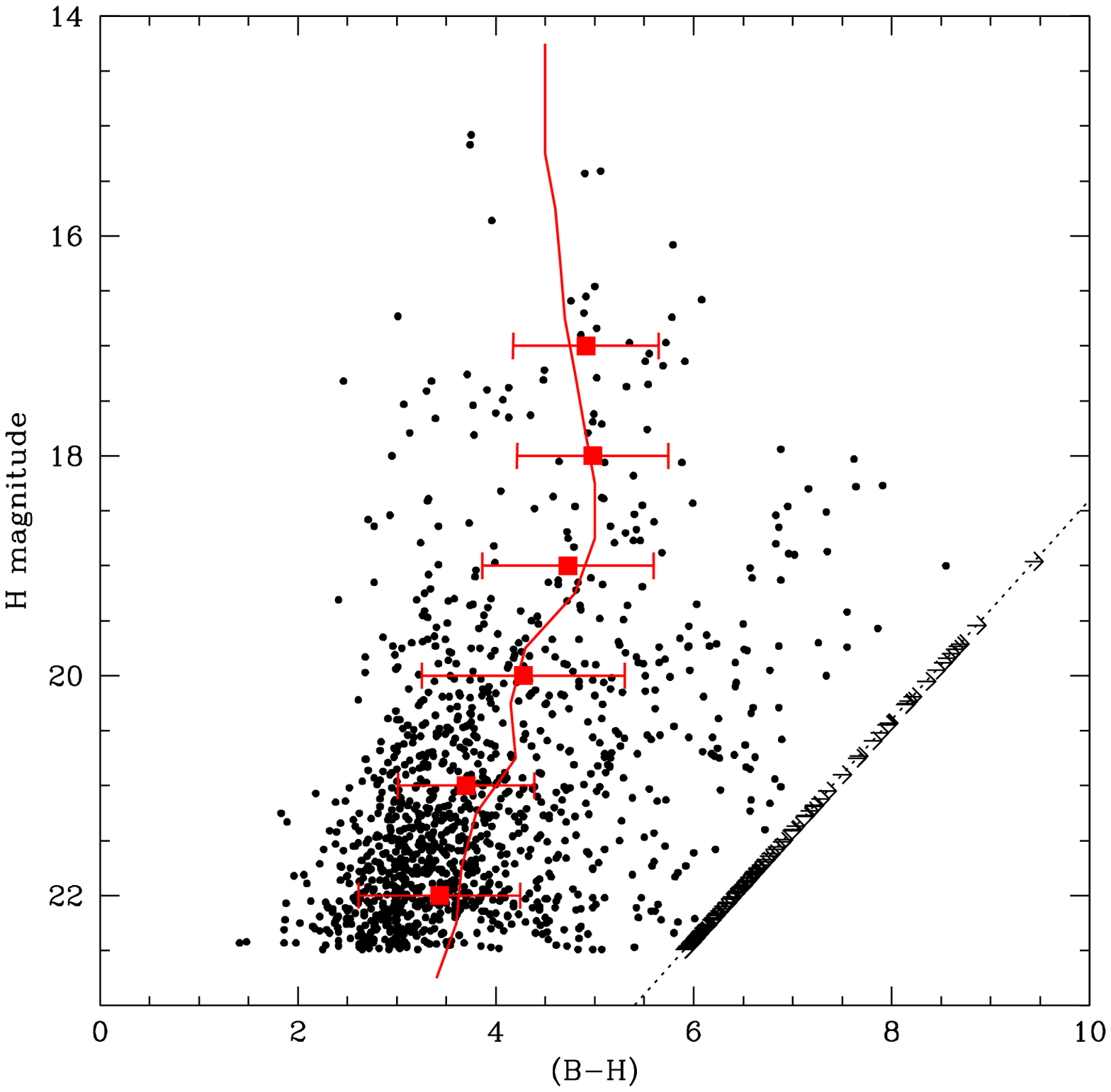}
\caption{$H$ magnitude against $B-H$ colour for galaxies in the WHDF,
         with median colours calculated in one magnitude bins. The median of our
         evolving $q_0=0.05$ $x=3$ model is also shown (solid line).  The
         dashed line shows the region of incompleteness.  Right-pointing
         arrows show objects too red to be detected in the optical band.}
\label{fig:medBH}
\includegraphics[width=\hsize]{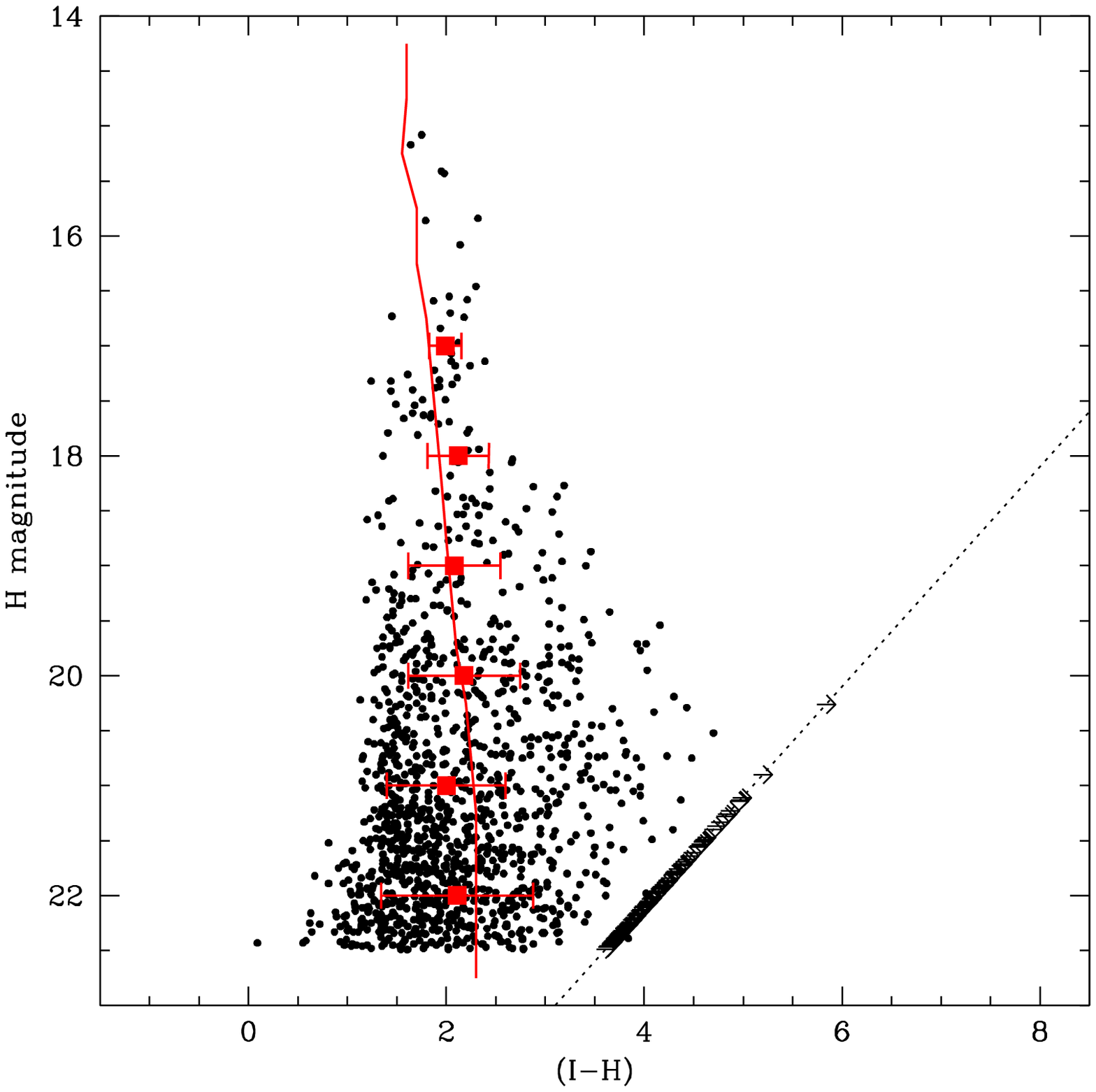}
\caption{As in Fig.~\ref{fig:medBH}, but instead $H$ magnitude against
         $I-H$ colour is plotted.}
\label{fig:medIH}
\end{center}
\end{figure}

Fig.~\ref{fig:medBH} and Fig.~\ref{fig:medIH} show optical - infra-red
colours for $H$-selected objects in the WHDF.  If compared to fig.~2
of Paper IV it is apparent from these graphs that the Calar Alto
dataset is a considerable advance over our old UKIRT wide survey,
which was limited at $K\sim20$. This work effectively extends our
$3\sigma$ limit two magnitudes fainter with the same area coverage and
with reduced photometric errors.  In Fig.~\ref{fig:medBH} we plot
$B-H$ colour against $H$ magnitude for all objects to $H<22.5$; also
shown are median colours (filled squares), and the predictions for the median
colours of the $q_0=0.05$ $x=3$ evolutionary model.  
Fig.~\ref{fig:medIH} is identical
to Fig.~\ref{fig:medBH} except that in this case we plot $I-H$ as a
function of $H$ magnitude. This can be compared directly with fig.~9
of \cite{CMM+02}.

In Fig.~\ref{fig:medBH} the median $B-H$ colour becomes slightly redder
to $H\sim18$, after which it turns bluewards and this trend continues to
the the limit of our survey. This should be compared to our previous $K$
vs $B-K$ plot (fig.~2 of Paper IV), which shows a similar trend for the
non-evolving models. There, however, a large apparent gap was seen in
the data near $K=21$, $B-K=5.8$ where very few galaxies were detected
although this was above the detection threshold in both filters. In the
present data this gap is not quite as obvious, although there is still a
blueward turn of the median of the median colour faintwards of
$H\approx20$.

\begin{figure}
\begin{center}
\includegraphics[width=\hsize]{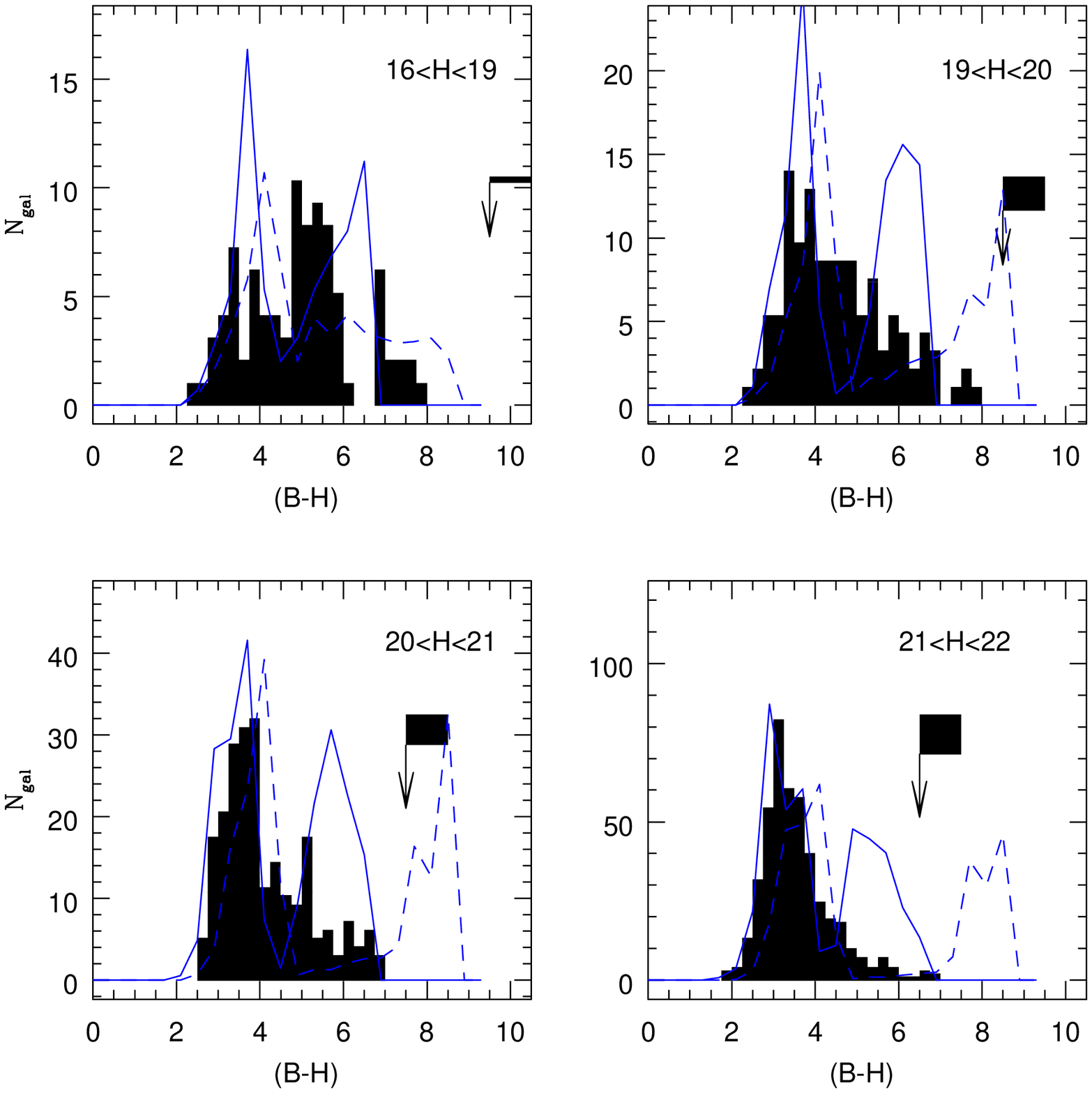}
\vspace{-0.25in}
\caption{$H$ selected $B-H$ distributions; arrows show
         the colour completeness limit -- the area of attached box shows
         the number of objects with no measured colour. The solid line
         show the predictions of the $q_0=0.05$ $x=3$ evolutionary
         model. The dashed line is the non-evolving counterpart.}
\label{fig:slicesBH}
\includegraphics[width=\hsize]{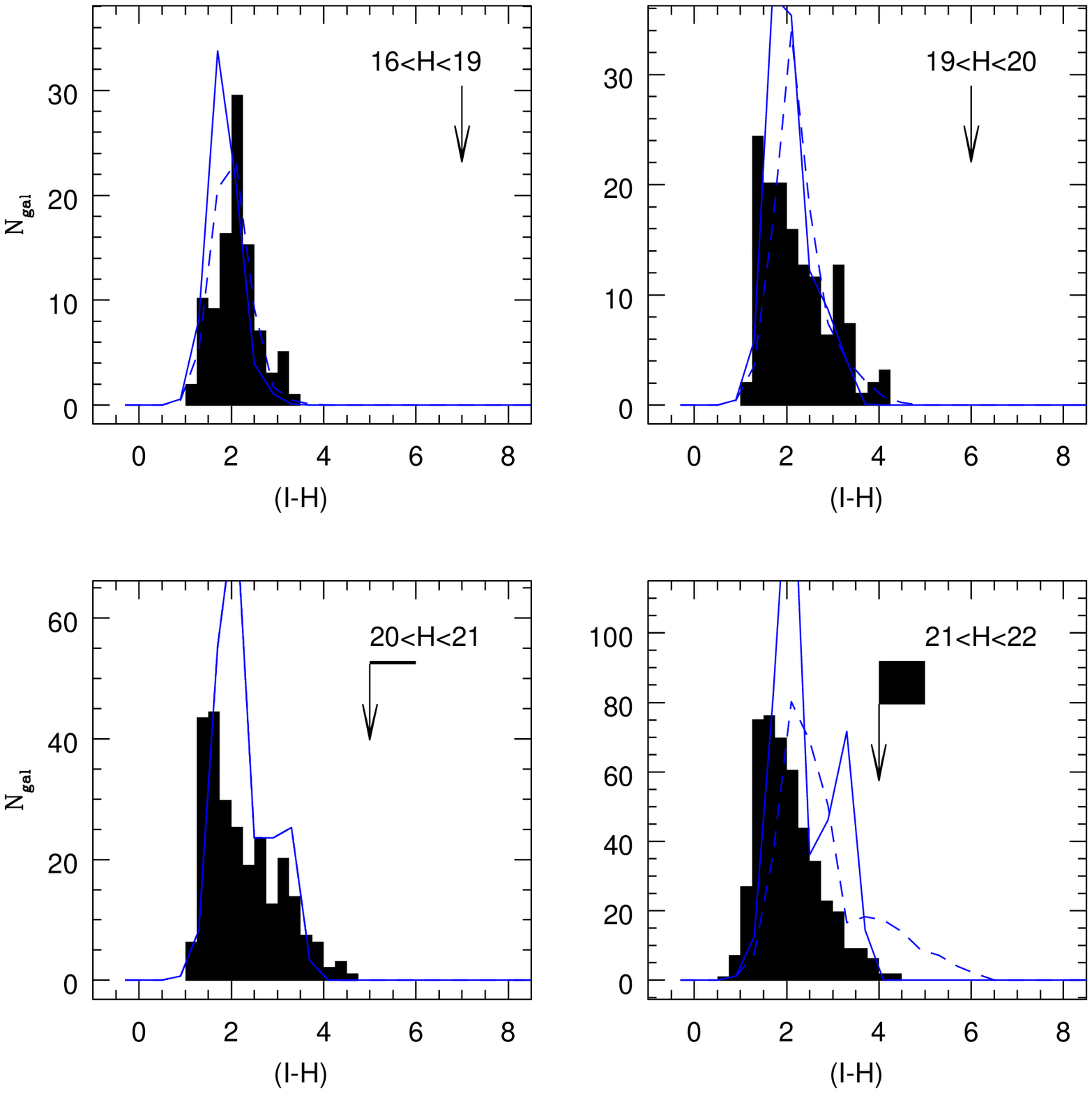}
\vspace{-0.25in}
\caption{$H$ selected $I-H$ distributions for the Calar Alto
         dataset. All symbols are identical to Fig.~\ref{fig:slicesBH}.}
\label{fig:slicesIH}
\end{center}
\end{figure}

The histograms of galaxy colours show this bluewards movement more
clearly. In Fig.~\ref{fig:slicesBH} and Fig.~\ref{fig:slicesIH} we
present the $B-H$ and $I-H$ colour distributions for objects in the WHDF
selected by $H$ magnitude in four slices from $16<H<19$ to $21<H<22$. 
The dashed lines show the predictions of a non-evolving model, whereas
the solid lines show the predictions from the $x=3$ evolutionary model.
We have {\it not} renormalised the models counts to agree with the data
in each bin, and as a result the evolutionary histogram slightly
overpredicts the numbers of objects.

\begin{figure}
\begin{center}
\includegraphics[width=\hsize]{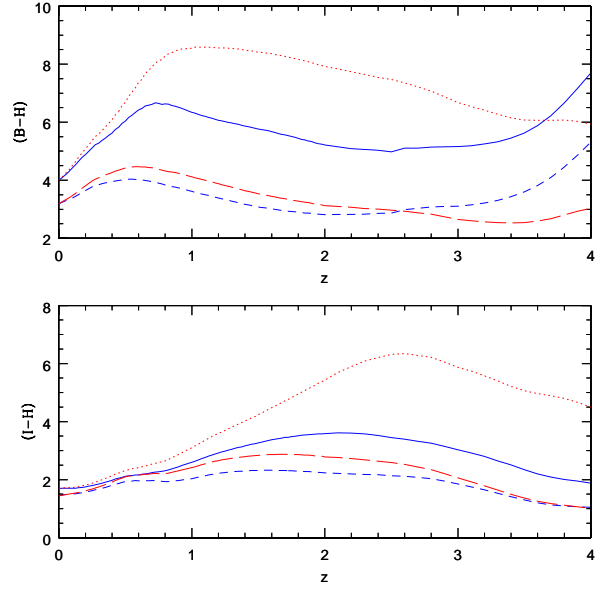}
\caption{$I-H$ colour (lower panel) and $B-H$ colour (upper panel)
         as a function of redshift for our $q_0=0.05$ model; non evolving
         E/S0 (dotted line), evolving E/S0 (solid line), non-evolving
         spiral (short-dash), evolving spiral (long-dash).}
\label{fig:cazed}
\end{center}
\end{figure}

These diagrams confirm the broad conclusions presented in Paper IV; we
find that the numbers of extremely red, unevolved objects present in
these distributions are extremely small.  In Fig.~\ref{fig:cazed} we
plot $I-H$ and $B-H$ colour of our models as a function of redshift.  
A non-evolving
galaxy track (i.e., a pure $k$-correction) reaches $I-H\sim6$ by
$z\sim2$, or $B-H\sim8$ by $z\sim1$. Our survey shows a conspicuous lack
of such objects. However, even with our $x=3$ model there is a red peak
predicted in the colour distributions at $B-H\sim6$ and, in the faintest
bin, at $I-H\sim3$  which is not seen in the data. These are the evolved
E/S0s at $z>1$, which although much bluer than non-evolving predictions,
still should be present in our sample.

Following, Paper IV, it is worth considering the $I-H$ histograms in
Fig.~\ref{fig:slicesIH} more carefully. While the brighter H bins
continue to show the slightly extended red tail that indicates the
continuing presence of the early-type galaxies, by the faintest
$21<H<22$ bin, there may be more of a case that these galaxies could
have disappeared (de-merged). However, these galaxies may have simply
moved bluewards faster than the PLE model as the UV flux enters the I
band at $z\approx1$, as appears to have happened already in the B band (see
Fig.~\ref{fig:slicesBH}) for galaxies 2mag brighter in H. 
Indeed, renormalising the model prediction
downwards suggests that the overall shape of predicted  $I-H$
distribution may still fit the data even in this faintest bin. Thus we
conclude that the early-type galaxy population may persist essentially
unchanged except in the UV out to $z>1$, as indicated by the continuing
excellent fit of the PLE models in $I-H$ to $H\approx22$ mag.

\begin{figure*}
\begin{center}
\includegraphics[width=0.8\hsize]{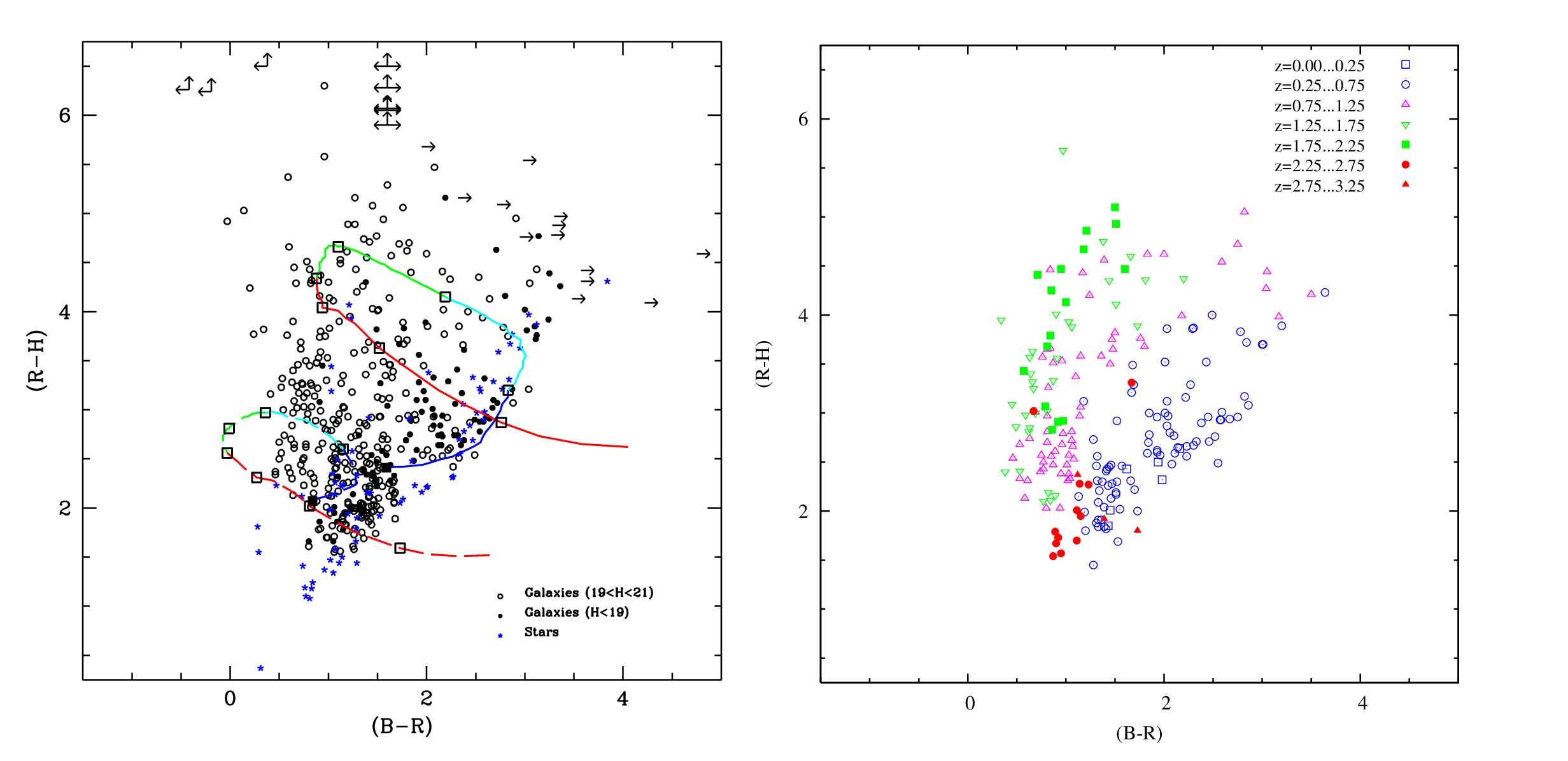}
\caption{Left: $H-K$ vs $R-H$ plot for all galaxies and stars on the
         WHDF with $H<21$. Filled and open circles represent galaxies
         brighter and fainter than $H=19$ respectively. Colour limits
         for those not detected in $R$ are shown with arrows, assuming
         a magnitude $0.5$\,mag below the $3\sigma$ $R$-band detection
         threshold. A double horizontal arrow indicates galaxies not
         detected in $B$ or $R$ (i.e.\ its $B-R$ colour is undetermined).
         Also shown are the Bruzual and Charlot tracks for E/S0 galaxies
         (solid line) and Scd galaxies (dashed line). Redshift intervals
         of $z=0,0.5,1,1.5,2,2.5,3,3.5$ are marked with boxes. Right:the
         same colour-colour plot, but only for those galaxies assigned
         photometric redshifts from Hyperz with $>80$ per cent probability.
         Redshift ranges are indicated by the different symbols.}
\label{fig:brrhplot}
\includegraphics[width=0.8\hsize]{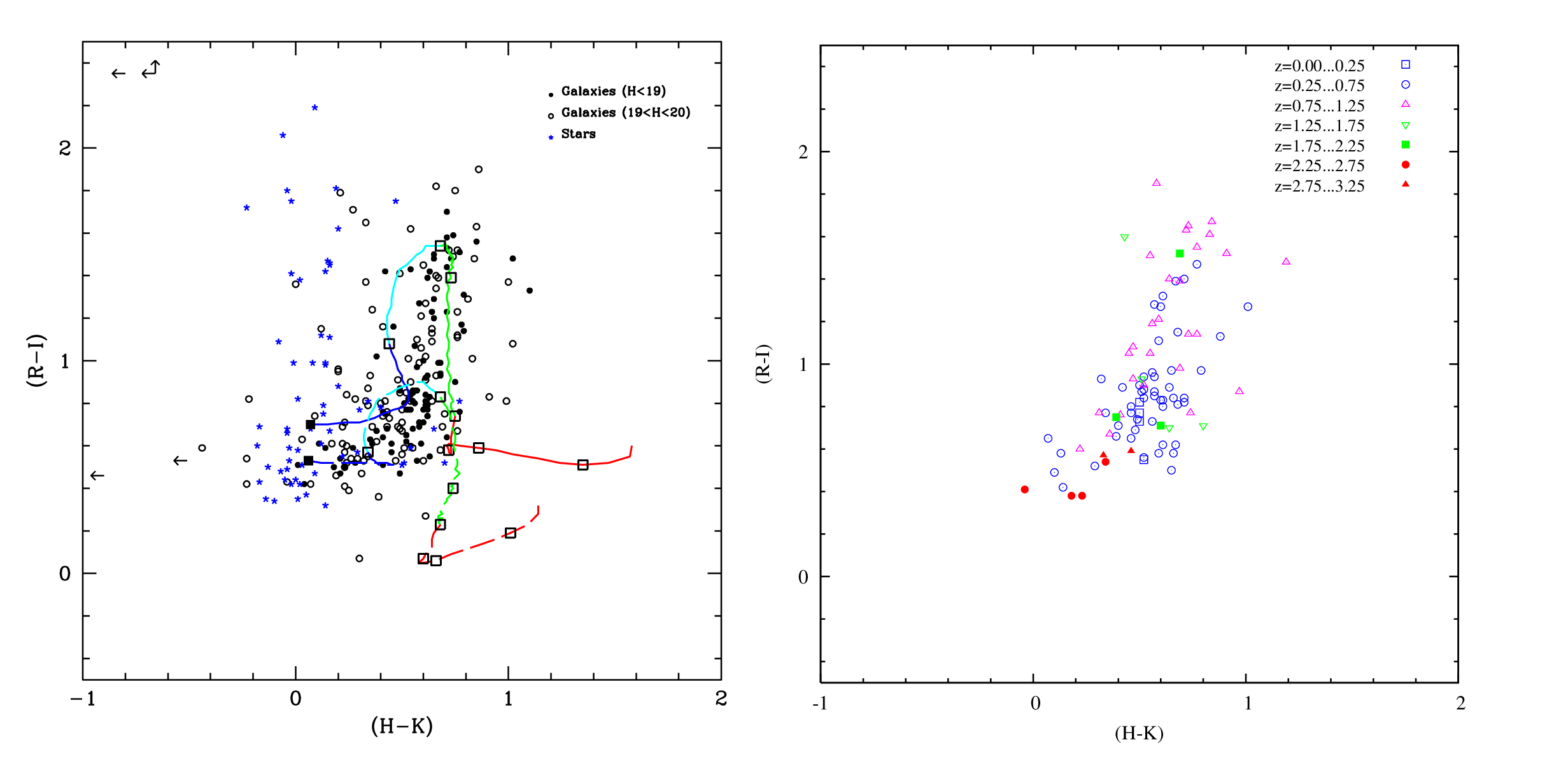}
\caption{As Fig \ref{fig:brrhplot} but now for $H-K$ vs $R-I$ and
         limited at $H=20$.}
\label{fig:hkriplot}
\end{center}
\end{figure*}

\begin{figure}
\begin{center}
\includegraphics[width=3.25in]{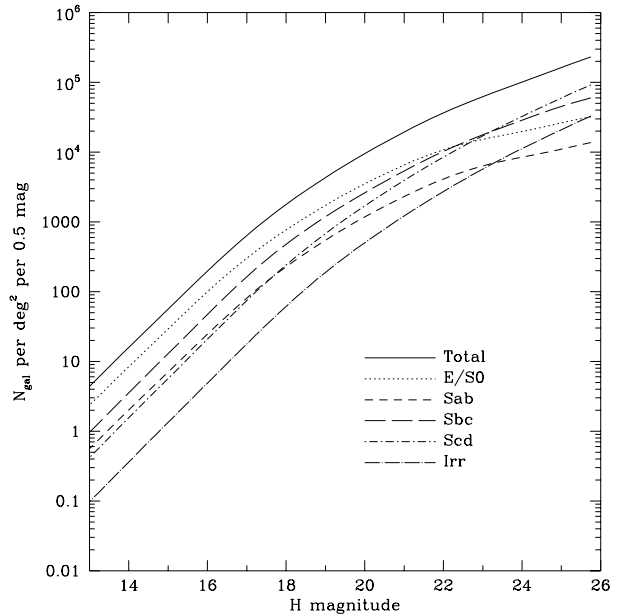}
\caption{$H$-band number-magnitude relation as a function of
  morphological type for the evolutionary $q_0=0.05$ model. The heavy
  solid line shows the total predicted galaxy count.}
\label{fig:types}
\end{center}
\end{figure}

Fig.~\ref{fig:brrhplot} shows $B-R$ vs $R-H$ for all galaxies in the
WHDF, as well as the colour tracks predicted by the models. In this
plot, galaxies have been split by magnitude, with the filled circles
representing galaxies with $H<19$. From this it is apparent that a large
number of the faintest galaxies lie in the region of this plot occupied
by spiral galaxies. This is not unexpected, as indicated by
Fig.~\ref{fig:types}, which shows the number counts from our model for
the individual morphological types.

As noted in Paper IV, and by \cite{CMM+02}, there is a large scatter in
the colour-colour plane, with galaxies distributed over a very broad
range of optical - infra-red colours, particularly for the objects the
models tracks suggest should be early types (remember that for clarity
we only show one spiral track, but in reality the Sbc-Irr tracks will
represent most of the bluer colours). The large scatter could explain
the flat distribution in Figs~\ref{fig:slicesBH} and \ref{fig:slicesIH}
compared to the models at $B-H\sim6$ and $I-H\sim3$. This is discussed
further in the next section, but may be evidence of a wide range of star
formation histories, dust content, or metallicities for the early types.
It is interesting to speculate whether the intermediate population of
`blue' early-types detected by Vallbe et al.\ (in prep.) at $z<0.5$ are now
contributing at higher redshifts to the wide scatter in $R-H$ at $z>0.5$.
The problem is that the scatter is as much on the red side of the track
as on the blue in this redshift range. However, the early-type track in
$R-H$ is quite sensitive to e-folding time of the SFR, $\tau$ (which is
degenerate with the IMF slope). It may be possible to choose
$\tau<2.5$Gyr in the $x=3$ case to make the track more like a red
envelope in the $B-R:R-H$ diagram at $0.5<z<1.5$ while not making the
track too red at $z\sim0.5$ in $B-R:R-I$. Then the scatter would shift to the
bluewards side and be explained by on going star-formation in the
intermediate early-type population.

Fig.~\ref{fig:hkriplot} shows $H-K$ vs $R-I$. This time we curtail
the distribution at $H=20$, due to the bright limit of the $K$ data. This
confirms the suggestion from the optical data in Paper V that the
galaxies redder than $R-I\sim1$ are likely to be early types in the range
$0.5<z<2$.

To illustrate these points further, we also carried out a comparison
of photometric redshifts of the $H$-band selected galaxies with our
elliptical and spiral models. The photometric redshifts were derived
using the Hyperz package \citep*{HyperZ}, publicly available at {\tt
http://webast.ast.obs-mip.fr/hyperz/}. As input to Hyperz we used our
catalogue of galaxies with $H<21$\,mag and a maximum of six magnitudes
in the filters $U$,$B$,$R$,$I$,$H$, and $K$. We let Hyperz compute the
most likely redshift in a range $0.1 < z < 6.0$ with steps of 0.05 and
a possible internal reddening $0.0 < A_V < 2.0$\,mag with the Calzetti
law \citep{Calz97}.  The full set of observed and model templates coming
with Hyperz was used to cover all observed types, but checks with limited
template samples showed no significant difference. To be consistent with
our previous results we used the cosmology with $H_0=50$, $q_0=0.05$
without a cosmological constant. The weighted mean redshifts with
corresponding confidence probability better than 80 per cent were selected. A
check with known redshifts in the WHDF suggests typical errors of
$\delta z = \pm0.10$ (for $z<0.5$) if the object is observed in 
all six filters. The
result is shown in the right-hand panels of Figs. \ref{fig:brrhplot}
and \ref{fig:hkriplot}.

In general, we find a good agreement between the our model tracks and
the photometric redshifts of our galaxy sample in these two-colour
diagrams. Our model tracks in the range $0.0 < z < 0.5$ are close to the
centre of the area covered by the objects with the photometric redshift
in this range as shown in the $B-R$ vs $R-H$ colour plane
(Fig.~\ref{fig:brrhplot}). Only a few $z>2.25$ objects are scattered in
the region where our models have redshifts below $z=2$. Note that the
location of the low redshift galaxies in this diagram is about the same
as that of the brighter $H<19$ galaxies is in Fig.~7. The galaxies with
bright apparent magnitudes are therefore mostly low redshift objects.

In the $H-K$ vs $R-I$ two-colour plane (Fig.~\ref{fig:hkriplot}) the
photometric redshifts of only the lowest redshifts ($z\la0.5$) agree
with the corresponding model tracks of Fig.~8. The objects in the range
$0.75 < z < 1.25$ are distributed over the whole diagram, whereas this
redshift range in the model tracks is located in a very small area. And
especially the $z>2$ objects are much bluer in $H-K$ than our model
tracks predict. This is due to the type selection of the Hyperz
templates: the best fit template of the $z>2$ objects is a starburst
seen at a young age, a case not included in our simple models.

\subsection{EROs}

\begin{table}
\caption{Numbers of EROs with $R-K>5$, $K<19.2$ per $\sq\arcmin$
         and as a percentage of all galaxies with $K<19.2$.}
\label{tab:kero}
\begin{center}
\begin{tabular}{lcc}
\hline
Author& $N_\mathrm{ERO} / \sq\arcmin$&Percentage\\[0.5em]
\hline
Model ($x=3$)&0.67&11\\
Model (Salpeter)&2.7&29\\[0.5em]
This work&$0.53\pm0.1$&$15\pm0.03$\\
\citet{Dadd00}&0.63&13\\
\citet{CDM+02}&0.88&$13\pm0.02$\\
\citet{Roch03}$^{\dag}$&0.5&13\\
\hline
\end{tabular}\\[0.5em]
$^{\dag}$ Estimated from values at $K=19$ and $K=19.5$
\end{center}
\end{table}

Much attention has been devoted in recent years to the study of Extremely
Red Objects, or EROs (e.g.~\citealt{CDM+02}, \citealt{Smit02}, \citealt{Roch03},
\citealt{Yan03}). These are objects traditionally selected to
have $R-K>5$, although as is clear from Fig.~\ref{fig:brrhplot} and
Fig.~3 of Paper IV, this has little meaning in the context of modern
evolutionary models other than to select E/S0s roughly in the range
$1<z<2$. Nevertheless, we show a comparison of our numbers for $R-K>5$
with other published values in Table \ref{tab:kero}. Although the
absolute numbers vary, the percentage of EROs is remarkably constant
between authors.  We also show the prediction of the $q_0=0.05$ $x=3$
model, and a standard Salpeter PLE model. Several problems affect
these comparisons; firstly the red extreme of the model tracks is very
sensitive to the exact star formation history adopted, and, secondly,
$R-K=5$ is on a steeply falling portion of the number-colour histogram,
so small uncertainties in zero-point and random errors on the colours
(and potentially colour equations between $K$-bands used at different
observatories) could make substantial differences to the numbers. The
surface densities found by \cite{Dadd00} differ by almost a factor of
two between samples with $R-K>5.0$ and $R-K>5.3$.  Nevertheless, the
$x=3$ model predicts both a similar percentage as the data and absolute
densities close to the mean observed density at $R-K>5.0$. Our Salpeter
PLE model is a factor of two higher, as expected from the results in
Sect.~\ref{sect:IMF} where it was shown that this model predicts far
too many objects above $z=1$. Remaining small differences in the $x=3$
model can be explained by the observational result of \cite{CDM+02}
who presented evidence that a large fraction of the ERO population are
dusty starburst galaxies. A similar conclusion was reached by \cite{Yan03},
who found from HST imaging that about 65 per cent of EROs were showed disks,
although \cite{Yan04} found that the fraction of emission line objects was 
similar amongst both bulge and disk domimated EROs.
 As dust does not have a large effect on the
numbers of NIR selected galaxies it would shift the galaxy
population towards redder colour, so that the inclusion of dust in our
models would result in a somewhat higher number of EROs predicted.  
As the deviation of
the modelled number of EROs from the mean observed number is smaller than
the deviation between different the number from observational projects,
it makes little sense to try and finetune this parameter.

This success of the $x=3$ models with our new data and other data from
the literature somewhat contrasts the analysis of \cite{Smit02}. They
observationally found an order of magnitude more EROs (colour
cut $R-K>5.3$) than predicted by our model from Paper IV, in contrast
to more specialised models of \cite{Dadd00}. Since our early-type
track does not reach $R-K=5.3$ at any redshift, it is clear that 
our simple model will underpredict the the numbers of EROs with a 
very red cut.

\begin{table}
\caption{Numbers of objects with $I-H>3$ per $\sq\arcmin$ with
         $\pm1\sigma$ errors over the $1.26\times10^{-2}$ deg$^2$ of
         the deep $H$ image which has $I-H$ colours. Objects too red
         to have a measured colour are included.}
\label{tab:hero}
\centering
\begin{tabular}{@{}l ccc@{}}
\hline
Author               & $19<H<20$     & $20<H<21$    & $21<H<22$    \\[0.5em]
\hline
Model (evol.)$^\ast$ & $0.3$ (8\%)   & $1.4$ (18\%) & $4.1$ (27\%) \\
Model (NE)           & $0.5$ (16\%)  & $1.7$ (27\%) & $3.4$ (29\%) \\[0.5em]
This work            & $0.55\pm0.1$  & $1.17\pm0.2$ & $2.1\pm0.2$  \\
                     & (18\%)        & (21\%)       &  (20\%)      \\
\cite{CMM+02}\hspace*{-1.5em}& $0.42$ (15\%) & $1.1$ (20\%) & $-$  \\
\hline
\end{tabular}\\[0.5em]
$^\ast$ Evolving model with $q_0=0.05$ and $x=3$ IMF
\end{table}

\begin{figure}
\begin{center}
\includegraphics[width=\hsize]{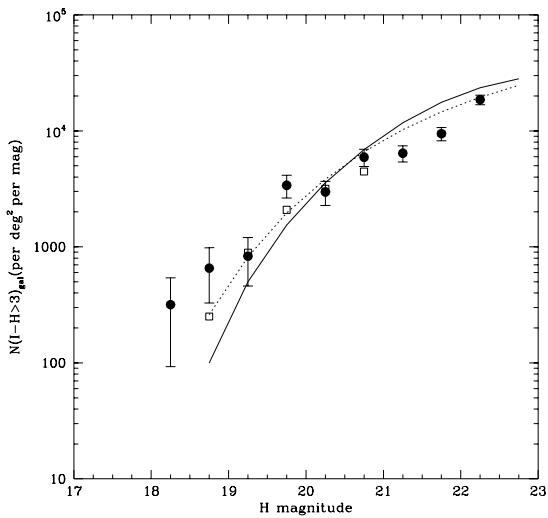}
\caption{The number-counts of EROs with $I-H>3$. Filled circles, this
         paper; open squares, \protect\cite{CMM+02}. Our faintest
         point is on the limit of the $I$-band photometry and should
         be treated with caution. The solid line shows the prediction
         of the $q_0=0.05$ $x=3$ model, the dashed line that of the
         non-evolving $q_0=0.05$ model.
}
\label{fig:eros}
\end{center}
\end{figure}

With our deeper $H$-band data we can study the distribution in $I-H$ as
previously done by \cite{CMM+02}. Table \ref{tab:hero} shows the numbers
of objects with $I-H>3$ -- a colour-cut that is roughly equivalent to
$R-K>5$, although it selects a slightly higher redshift range -- for
three $H$ magnitude bins. For the bins in common, we find excellent
agreement between our data and that of Chen et al.. We show the
$H$-selected ERO counts subdivided into smaller $0.5$\,mag bins in
Fig.~\ref{fig:eros}.  While Chen et al.\ suggested that the number
counts turn over at $H\sim20$ it is clear from our deeper $H$-band data
that this does not happen. Instead, the counts continue to rise towards
our faintest bin centred $H=22.25$\,mag. 

When comparing our models to these data we find that on average the
numbers are in good agreement with the observed numbers of ERO galaxies.
This is true for both our no-evolution model and the evolving $x=3$
model.  For $21<H<22$ the models predict 27--29 per cent $I-H>3$ objects,
compared with 20 per cent seen in the data. As indicated by Fig. 11, our ERO count
predictions using $B-H$ limits will be in much less good agreement
with the data, although as pointed out in Paper IV, due to the
sensitivity of the $B$-band to evolution, these colours can be changed
quite drastically by small changes in IMF or star-formation rate
e-folding time.

The location of our EROs with $H<21$ in four colour-colour planes are shown 
in Fig.~\ref{fig:es0cols}, together with our early-type model track. 
The data show a large scatter (of over 1 magnitude), 
but in all the plots tend to cluster around
the $z\sim2$ model location. Nearly half our EROs are detected in the $U$ band,
but this should not be taken as evidence of unusual star forming activity, as
the model colour for an E/S0 at $z\sim2$ is $U-H\sim5.5$, well within the
range of our $U$ data for $H<21$.

\subsection{The Early-type Sequence}

\begin{figure*}
\begin{center}
\includegraphics[width=5in]{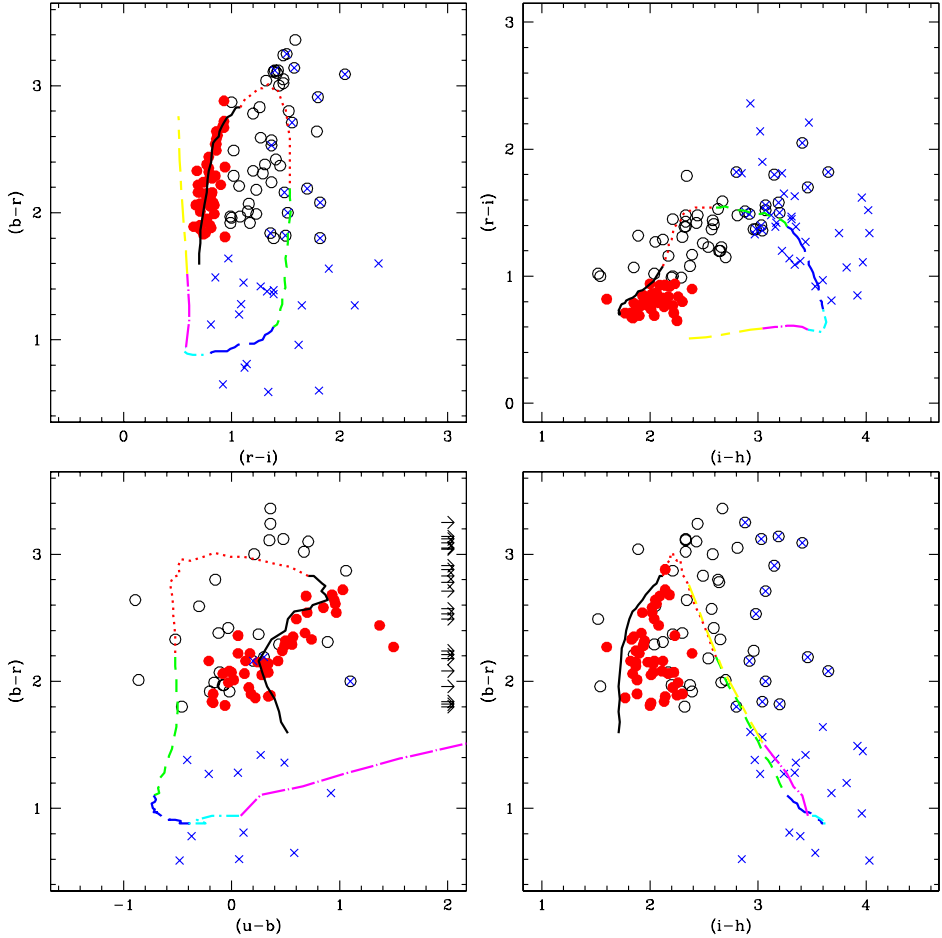}
\caption{The colour properties of E/S0 'sequence' galaxies.  Shown are
         four colour-colour diagrams (all limited at $H<21$) for galaxies
         selected on the basis of their location in the $R-I$:$B-R$
         plane (top left). Solid circles, $B-R>1.8$, $R-I<0.95$
         -- these are identified as E/S0 galaxies with $z<0.5$; open
         circles, $B-R>1.8$, $R-I>0.95$ -- potential E/S0 galaxies
         in the range $0.5<z<1.0$ . Also shown, as crosses are the
         location of EROs defined as $R-K>5$.  Our $q_0=0.05$ $x=3$
         evolving model for E/S0s is shown; solid line, $0<z<0.5$;
         dotted line, $0.5<z<1.0$; short dashed line, $1.0<z<1.5$;
         long dashed line, $1.5<z<2.0$; short dash-dot, $2.0<z<2.5$;
         long dash-dot, $2.5<z<3.0$; long dash short dash, $3.0<z<3.5$.}
\label{fig:es0cols}
\end{center}
\end{figure*}

Previous studies (with the exception of \citealt{FSM+02}) have tended to
concentrate simply on one colour. Our multi-wavelength data enables us
to explore galaxies with particular properties in different two-colour
planes. In particular, in Paper IV we showed how the $R-I$:$B-R$ plane 
\emph{seems} to provide a means of
selecting early-type galaxies. According to the models, the clear
'sequence' seen with $B-R>1.8$ and $R-I<0.95$ delineates E/S0 galaxies
with $z<0.5$.  In fact this is 
insensitive to choice of model --
a simple $k$-correction would come up with a similar cut. The models
also suggest that $B-R>1.8$ and $R-I>0.95$ should select E/S0s between
$0.5<z<1$.  The $H$-band is ideal to study if this truly selects early
type galaxies.  In Fig.~\ref{fig:types} we plot the contribution of
the various morphological types of galaxy to the $H$-band number-counts
for our $q_0=0.05$ $x=3$ model. It can be seen that early type galaxies are
the most numerous type up to $H\sim22$.  Fig.~\ref{fig:es0cols} shows
the location of our colour selected E/S0s with $H<21$ (this ensures good
colour completeness in all except $U-B$) in four colour-colour planes,
together with the redshift coded model tracks for E/S0s.  The reader's
attention is drawn to the fact that many of the low $z$ galaxies at
$B-R\sim2$ appear too blue in $U-B$ and too red in $I-H$ to be normal
E/S0s, suggesting that these galaxies have experienced more recent
star-formation than contained in our simple PLE model.
This may be  evidence for an intermediate population of early-type
galaxies. Furthermore, some of these galaxies, although lying on the
early-type sequence in $R-I$:$B-R$, have been found to have bluer $B-R$ colours
than expected for their redshift, again suggestive of an intermediate age
early-type population. This is discussed by Vallbe et al in prep where
evidence for an intermediate early-type population in a 2dFGRS/SDSS
dataset is also investigated.

 Recently, new modelling techniques that include the thermally pulsing
 asymptotic giant branch (TP-AGB) of some stars were suggested
 \citep[see e.g.][]{SFA+02} that find much redder optical-NIR colours
 for stellar populations of ages $>10^8$\,yr, a feature that was
 successfully tested on globular clusters \citep{MKB+01} and distant
 galaxies \citep{SFA03}. This property is not included in our models
 based on GISSEL99 \citep{BC93}, but a combination of this feature and
 moderate amounts of dust may well produce a UV excess plus extra
 optical-NIR reddening in intermediate age stellar populations, as seen for the 
 early-type galaxies in Fig.~\ref{fig:es0cols}. The
 spectral properties of these galaxies will be discussed in a subsequent
 paper where we will also explore the effect of refinements of our
 modelling technique including the TP-AGB.

\subsection{A new high-$z$ galaxy cluster}
\begin{figure*}
\begin{center}
\includegraphics[width=\hsize]{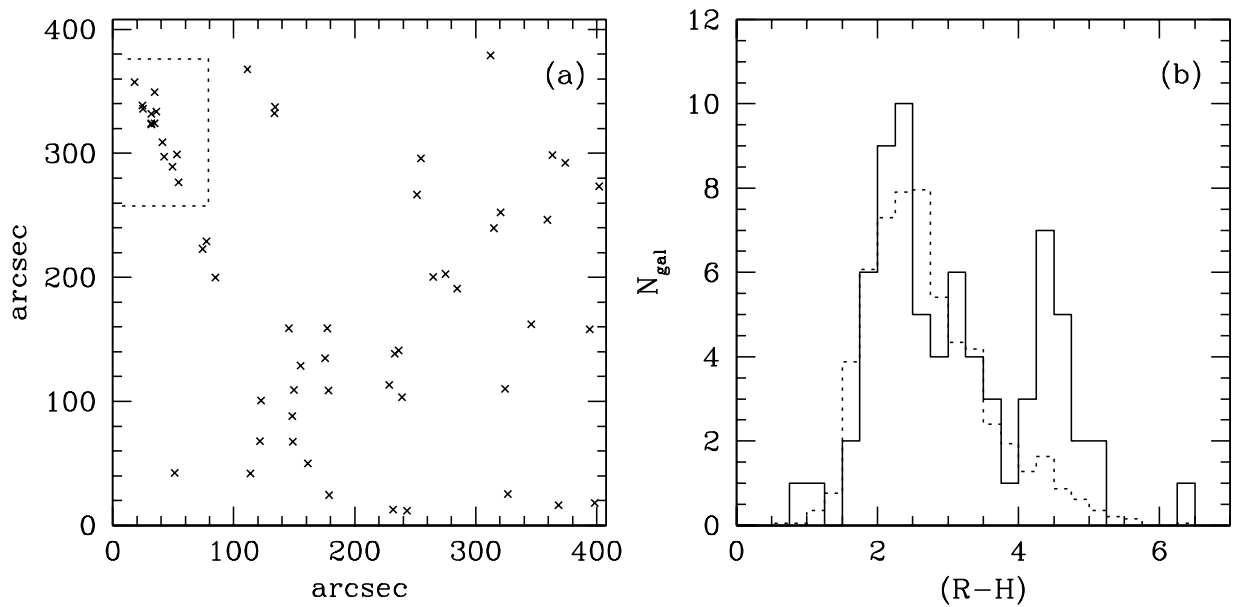}
\caption{{\bf (a)} The position of galaxies with $I-H>2.5$, $R-I>1$
         over the $7'\times7'$ field of view of the WHDF. The region
         containing a possible new high redshift galaxy cluster is
         highlighted.
         {\bf (b)} Distribution of $R-H$ colours for galaxies in
         a 2.35\,$\sq\arcmin$ region centred on our proposed cluster
         (solid histogram), compared with the distribution for the whole
         field, normalised to the same area (dashed histogram).}
\label{fig:cluster}
\end{center}
\end{figure*}

Our colour selection has identified a possible high redshift cluster on
the WHDF.  Selecting galaxies with colours $I-H>2.5$ and $R-I>1$
and plotting their location on the sky, shows a pronounced overdensity
of some 20--25 galaxies of objects within a diameter of $1.5'$ ($\sim$
0.5$h^{-1}$\,Mpc at $z\sim1$) in a region near the NW corner ($00^h
22^m 45.2^s$, $0\degr23'26''$) of our $7'\times7'$ field of view.
This region is highlighted in Fig.~\ref{fig:cluster}a. A histogram of
galaxy $R-H$ colours within this region in Fig.~\ref{fig:cluster}b
(solid line) shows a second peak at $R-H\sim4.5$ while there is
no extra peak in galaxy colours over the whole field (dotted line).
Using our optical and NIR magnitudes of the possible members
of this cluster we have computed photometric redshifts with Hyperz and
find that most of these galaxies are between $1<z<1.5$, most likely in
the lower half of this redshift range.  The brightest three galaxies in
the region have $H\sim19$, although the one nearest the apparent centre
of the concentration has very blue colours, which would imply a lower
redshift late-type galaxy.

Of course, these are only indications for the discovery of a new
galaxy cluster at $z\gtrsim1$, which would require spectroscopic follow-up
for confirmation. But even so one might speculate about the apparent
shape of the overdensity which does not seem to be spherical but more
elongated. Although the cluster is detected near the edge of our field
of view and we should be careful not to overinterpret this, we might be
looking at a cluster in formation where galaxies from the surrounding
filaments are falling into the gravitational potential of the cluster.

%%%%%%%%%%%%%%%%%%%%%%%%%%%%%%%%%%%%%%%%%%%%%%%%%%%%%%%%%%%%%%%%%%%%%%%%%%%%%%%%

\section{Conclusions}\label{sect:conc}
We have presented data from deep NIR observations of the
William Herschel Deep Field down to $H\sim22.5$\,mag. These data reach
about two magnitudes deeper than previous ``wide area'' observations in
the NIR (see Paper IV) and now extend over the full central $7'\times7'$
of the WHDF. Several conclusions can be drawn from these new data and in
comparison with our models. These PLE models assume a bimodality in SF
history; red galaxies essentially evolve passively after an initial
burst of star-formation at high redshift whereas blue galaxies evolve
with star-formation only decaying with an e-folding time of 9\,Gyr.

In our investigation about input parameters for our PLE models we
noticed that a discrepancy exists between galaxy luminosity functions
that were derived in the optical and NIR wavelength ranges.
Specifically, LFs derived from 2MASS data seem to have a much shallower
faint end slope than the LFs derived from 2dFGRS data in the $b_J$-band.
We also confirm the observation of the K20 redshift survey team that
galaxy redshift distributions $N(z)$ of models using the standard
Salpeter IMF predict too many high redshift objects while models using
IMFs with steeper slopes like the Scalo IMF or our favoured $x=3$ IMF
very well match the observed $N(z)$ in the $K$-band.

The models also give reasonable fits to  the early-type LF's in the rest
$B$-band out to $z\approx0.8$ from the photo-z COMBO-17 survey
\citep{BWM+03} and more exact fits to early results from the SDSS-2dF
Luminous Red Galaxy Redshift Survey out to $z\approx0.7$ (D.\ Wake, priv.\ 
comm.). Preliminary results from the  VVDS \citep{LeF04} continue to
show little evolution for the red galaxy LF and  out to z$\sim1$ but
about 1 mag luminosity evolution in the rest $B$ band out to $z\approx1$,
as expected from our simple PLE models. It is then interesting to check
whether these models continue to fit our galaxy counts and colours to
our faint $H$ limits.

Given the success of the models with $x=3$ or even a Scalo IMF, we note
that this would mean virtually no evolution in stellar mass over large
look-back times for early-type galaxies. Since conversions to stellar
mass from NIR luminosity are heavily dependent on the assumed IMF, in
this paper we have preferred to discuss evolutionary models in terms of
the early-type galaxy LF, which is the basic observed quantity, rather
than its stellar mass  derivative.

Taking the $H$ galaxy number counts first, we confirm most results noted
in Paper IV but now the data extends to $H\sim22.5$\,mag over a
7$'\times$7$'$ area and to $H\sim29$\,mag in the HDF-N$+$S. Models in
cosmologies with a high density parameter (i.e.\ $q_0=0.5$)  generally
underpredict the data. In the optical bands, these models were
supplemented with an additional population of early type dwarf galaxies
to address this problem; in the NIR these models slightly
over-predict the H counts at the faintest limits. Both evolving and
non-evolving models in cosmologies close to the so-called
``concordance'' parameters (i.e.\ with $\Omega_\Lambda = 0.7$) or
cosmologies with low $q_0=0.05$ give good agreement to the observed H
number counts to the faintest limits. In these cases, models that assume
our original `steep' $H$-band LF locally  give much better fits to the
faintest counts than the recent flatter LFs from 2MASS \citep{CNB+01}. 
These latter models tend to under-predict the numbers of galaxies
at $H>21$\,mag. If the steep local $H$-band LF is correct then the
suggestion is that the form of the galaxy LF at $z\sim$1--2 has not
evolved since the present day. If the flat LF is correct then the galaxy
LF at $z\sim$1--2 has steepened significantly with look-back time.

In terms of galaxy colours, we continue to find a deficiency of very red
galaxies at $H>20$\,mag. Median colours per magnitude bin are reasonably well
fitted by PLE models but only poorly reflect the distribution in colour in
each magnitude bin. We have demonstrated this using optical$-$NIR colour
histograms, where both evolving and non-evolving models predict more red
galaxies than are detected at $H>20$\,mag. Since the $H$-band counts are
well fitted by the models and since the effect is smaller in $I-H$ than
$B-H$, the models may somewhat underestimate the evolution in the $B$-band
at low redshift and in the $I$-band at higher redshift. This effect may
be related to the existence of an intermediate early-type population;
despite appearing tightly tied to the early-type locus in $B-R$:$R-I$,
the $B-R$ colours are frequently too blue for an early-type galaxy at a
given redshift. This intermediate early-type
population is also seen at low redshift in the 2dFGRS data 
(Vallbe et al.\ in prep.).
The intermediate population may comprise 30 per cent of the
early-types on the $B-R$:$R-I$ track; these galaxies may have experienced
a burst of star-formation at relatively recent times and their existence
means that the bimodality in star-formation histories may not be exact.

The colour spread of early-type galaxies in two-colour diagrams that
involve an optical-NIR colour is larger than seen in diagrams that only
involve optical colours. This was somewhat unexpected and not easily
explained by our simple model.  The spread  is most likely caused by
starbursts and dust and the tightness of the tracks in the optical bands
may be enhanced by optical selection. The intermediate population
detected by Vallbe et al.\ (in prep) that shows a blue excess in the 
optical bands
may also show a NIR excess in the red bands. At higher redshift this
intermediate population may explain the increased scatter seen in the
colour-colour diagrams of NIR selected samples.

Number counts of galaxy subsamples like extremely red objects (EROs)
agree very well with data from the literature and are reasonably well
matched by our models. For a detailed comparison of all the features
of ERO numbers and number counts the models will very likely have to be
refined to include dust and starbursts, as well as up to date stellar
isochrones. To carry out a comprehensive comparison, however, much more
data is needed as the deviations between models and data are of the same
order as the deviations between different datasets.

In addition to these results we presented evidence for the discovery
of a new galaxy cluster that might be observed in formation at $z\sim1$.

Finally, we emphasize that the counts and $N(z)$ distributions seen in
NIR selected samples continue to be well fitted by models which assume
virtually no evolution in the $H$-band LF. In hierarchical models such
as the standard $\Lambda$CDM model, the red population is expected to
show significant dynamical and luminosity evolution. Also the rate of
dynamical evolution is expected to vary with bulge halo mass. Since the
observed galaxy counts and number redshift relations show virtually no
evidence of evolution in the early-types at any luminosity, it will be
interesting to see if the semi-analytic models of galaxy formation can
arrange for the expected dynamical and luminosity evolution to conspire
to leave the early-type LF looking unevolved over virtually its whole
luminosity range.

%%%%%%%%%%%%%%%%%%%%%%%%%%%%%%%%%%%%%%%%%%%%%%%%%%%%%%%%%%%%%%%%%%%%%%%%%%%%%%%%

\section*{Acknowledgements}
Based on observations collected at the Centro Astron\'{o}mico 
Hispano Alem\'{a}n 
(CAHA) at Calar Alto, operated jointly by the Max-Planck Institut f\"{u}r
Astronomie and the Instituto de Astrof\'{i}sica de Andaluc\'{i}a (CSIC).
HJMCC and NM acknowledge financial support from PPARC. PMW acknowledges
funding from European Commission through the ``SISCO'' RTN, contract
HPRN-CT-2002-00316.
The INT and WHT are operated on the island of La Palma by the Isaac
Newton Group at the Spanish Observatorio del Roque de los Muchachos of
the Instituto de Astrof\'{i}sica de Canarias.
Data reduction facilities were provided by the UK STARLINK project. We would
like to thank Ana Campos for assisting with the observations at Calar Alto.

%%%%%%%%%%%%%%%%%%%%%%%%%%%%%%%%%%%%%%%%%%%%%%%%%%%%%%%%%%%%%%%%%%%%%%%%%%%%%%%%

\bibliography{papervi.bbl}

\label{lastpage}

\end{document}